\documentclass[useAMS,usenatbib]{mn2e}
\usepackage{float}

\usepackage[figuresright]{rotating}
\usepackage{caption}
\usepackage{graphicx}

\usepackage{color}

\usepackage{natbib}	
\usepackage{url}

\usepackage{savesym}
\usepackage{amsmath}
\savesymbol{iint}
\usepackage{txfonts}
\restoresymbol{TXF}{iint}

\usepackage{subfigure}
\usepackage{epsfig}
\usepackage{enumerate}
\usepackage{url}
\def\simless{\mathbin{\lower 3pt\hbox
{$\rlap{\raise 5pt\hbox{$\char'074$}}\mathchar"7218$}}}   
\def\simmore{\mathbin{\lower 3pt\hbox
{$\rlap{\raise 5pt\hbox{$\char'076$}}\mathchar"7218$}}}   
\def\Msun{{\rm M}_\odot}                                       
\newcommand{\be}{\begin{equation}}
\newcommand{\ee}{\end{equation}}
\def\simlt{\mathrel{\hbox{\rlap{\hbox{\lower4pt\hbox{$\sim$}}}\hbox{$<$}}}}
\def\simgt{\mathrel{\hbox{\rlap{\hbox{\lower4pt\hbox{$\sim$}}}\hbox{$>$}}}}
\def\lesssim{\mathrel{\hbox{\rlap{\hbox{\lower4pt\hbox{$\sim$}}}\hbox{$<$}}}}
\def\gtrsim{\mathrel{\hbox{\rlap{\hbox{\lower4pt\hbox{$\sim$}}}\hbox{$>$}}}}
\def\v{\text{v}}
\newcommand       \bea          {\begin{eqnarray}}
\newcommand       \eea          {\end{eqnarray}}

\def\v{\text{v}}

\usepackage{color,soul}
\soulregister\cite7
\soulregister\citep7
\soulregister\ref7
\soulregister\pageref7

\usepackage{xcolor}

\title[Shock-powered V5589~Sgr]{Shock-powered radio emission from V5589 {Sagittarii} (Nova Sgr 2012 \#1)}   
\author[J. Weston et al.]{
Jennifer H. S. Weston,$^1$
J. L. Sokoloski,$^1$ 
Laura Chomiuk,$^{2}$ 
Justin D. Linford,$^2$ 
 \newauthor
Thomas Nelson,$^3$
Koji Mukai,$^{4,5}$ 
Tom Finzell,$^2$ 
Amy Mioduszewski,$^6$  
Michael P. Rupen,$^{7}$ 
\newauthor
 Frederick M. Walter$^8$ 
\\
$^{1}$Columbia Astrophysics Laboratory, Columbia University, New York, NY 10027, USA\\ 
$^{2}$Department of Physics and Astronomy, Michigan State University, East Lansing, MI 48824, USA\\
$^{3}$School of Physics and Astronomy, University of Minnesota, 116 Church Street SE, Minneapolis, MN 55455, USA\\
$^{4}$CRESST and X-ray Astrophysics Laboratory, NASA/GSFC, Greenbelt, MD 20771, USA\\
$^{5}$Department of Physics, University of Maryland, Baltimore County, 1000 Hilltop Circle, Baltimore, MD 21250, USA\\
$^{6}$ National Radio Astronomy Observatory, P.O. Box O, Socorro, NM
87801, USA \\
$^{7}$ National Research Council of Canada, Herzberg Astronomy and Astrophysics Programs, Dominion Radio Astrophysical Observatory, \\
	 P.O. Box 248, Penticton, BC V2A 6J9, Canada\\
$^{8}$ Department of Physics and Astronomy, Stony Brook University, Stony Brook, NY 11794, USA\\
}

\begin{document}

\maketitle
\label{firstpage}

\begin{abstract}
\noindent


Since the {\it Fermi} discovery of $\gamma$-rays from novae, one of the biggest questions in the field has been how novae generate such high-energy emission. Shocks must be a fundamental ingredient. Six months of radio observations of the 2012 nova V5589~Sgr with the VLA and 15 weeks of X-ray observations with {\it Swift}/XRT show that the radio emission consisted of: 1) a shock-powered, non-thermal flare; and 2) weak thermal emission from $10^{-5}$ M$_\odot$ of freely expanding, photoionized ejecta. Absorption features in the optical spectrum and the peak optical brightness suggest that V5589~Sgr lies 4~kpc {away} (3.2-4.6 kpc). The shock-powered flare {dominated} the radio light curve at low frequencies before day~100. The spectral evolution of the {radio} flare, its high radio brightness temperature, the presence of unusually hard ($kT_x > 33$~keV) X-rays, and the ratio of radio to X-ray flux near {radio maximum} all support the conclusions that the flare {was} shock-powered and non-thermal. Unlike {most} other novae with strong shock-powered radio emission, V5589~Sgr is not embedded in the wind of a red-giant companion. Based on the similar inclinations and optical line profiles of V5589~Sgr and V959~Mon, we propose that shocks in V5589~Sgr formed from collisions between a slow flow with an equatorial density enhancement and a subsequent faster flow. We speculate that the relatively high speed and low mass of the ejecta led to {the} unusual radio emission from V5589~Sgr, and perhaps also to the non-detection of $\gamma$-rays.

\end{abstract}

\begin{keywords} 
novae, cataclysmic variables --  binaries: general -- stars: variables: general -- stars: winds, outflows	 -- radio continuum: stars -- X-rays: stars
\end{keywords}

\section{Introduction}
\indent

\begin{table*}
 \centering

 \begin{minipage}{135mm}
\caption{Observations of V5589~Sgr in the radio and sub-millimetre. All observations were taken with the VLA, except those under program ID 2012A-S016, which were taken with the SMA.}
  \label{tab:Observations}
  \begin{tabular}{@{}lccccclc@{}}
  \hline
   Observation Date     & MJD  & Day$^a$ & Epoch     &  Program ID &Configuration&Observed Bands& Time$^b$(min)    \\
  \hline
2012 Apr 23.4 	 &	56040.4	&	2.4	  &	1 &	 	11B-170	 	&	C	 &	Ka	 &	23.9	\\
2012 May 22.0  &	56069.0	&	31.0	  &	A &		2012A-S016 	&	SMA (compact)	  &	 225 GHz	 &	60	\\
2012 Jun 6.2	 &	56084.4	&	46.4	  &	2 &		12A-479 		&	B	 &	C	 &	12.0	\\
2012 Jun 14.2 &	56092.2	&	54.2	  &	3 &		12A-479	 	&	B	 &  {Ka Ku X}&		19.5	\\
2012 Jun 22.3 &	56100.3	&	62.3	  &	4 &		12A-479	 	&	B	 &	X C S	 &	30.4	\\
2012 Jul 8.9	 &	56117.9	&	79.9	  &	B &		2012A-S016 	&	SMA (compact)	 &	 225 GHz	 &	208	\\
2012 Jul 11.1	 &	56119.1	&	81.1	  &	5 &		12A-479	 	&	B	 &	C L &		32.4	\\
2012 Jul 20.3	 &	56128.3	&	90.3	  &	6 &		12A-479	 	&	B	 &	{Ka Ku X}	 &	19.1	\\
2012 Aug 9.0	 &	56148.0	&	110.0 &	7 &		12A-479	 	&	B	 &	X C L	 &	30.4	\\
2012 Aug 11.2	 &	56150.2	&	112.2 &	8a &		12A-479	 	&	B	 &	{Ka Ku X} &		19.1	\\
2012 Aug 12.3	 &	56151.3	&	113.3 &	8b &		12A-479	 	&	B	 &	X C L	 &	29.9	\\
2012 Aug 28.0	 &	56167.0	&	129.0 &	9a &		S4322*	 	&	B	 &	C L		 &	17.7	\\
2012 Aug 28.0	 &	56167.0	&	129.0 &	9b &		12A-479	 	&	B	 &	X C L &		30.4	\\
2012 Aug 28.1	 &	56167.1	&	129.2 &	9c &		12A-483	 	&	B	 &	C L &		19.9	\\
2012 Sep 4.2	 &	56174.2	&	137.7 &	10 &		12A-483	 	&	B	 &	{Ka Ku X}	 &	19.1	\\
2012 Sep 7.9	 &	56178.0	&	140.0 &	11 &		S4322*	 	&	BnA   &	C L	 	&	17.7	\\
2012 Oct 31.0	 &	56231.0	&	193.0 &	12 &		12A-479	 	&	A	 &	{Ka Ku X}	 &	19.1	\\
2012 Dec 9.8	 &	56270.8	&	232.8 &	13 &		12A-479	 	&	A	 &	X C L &		30.4	\\
2012 Dec 23.6	 &	56284.6	&	246.6&	14 &		12A-479	 	&	A	 &	{Ka Ku X} &		19.5	\\
2013 Jan 5.7	 &	56297.7	&	259.8 &	15 &		12A-479	 	&	A	 &	X C L &		30.4	\\
2013 Feb 24.7	 &	56347.7	&	309.7 &	16 &		13A-461	 	&	D	 &	X C L	 &	29.9	\\
2013 Mar 7.6	 &	56358.6	&	320.6 &	17a &	13A-461	 	&	D	 &	{Ka Ku X}	 &	19.5	\\
2013 Mar 8.6	 &	56359.6	&	321.6 &	17b &	13A-461	 	&	D	 &	{Ka Ku X}	 &	19.1	\\
2013 May 29.4	 &	56441.4	&	403.4 &	18a &	13A-461	 	&	DnC-C &	X C L	 &	29.9	\\
2013 Jun 3.2	 &	56446.2	&	408.2 &	18b &	13A-461	 	&	DnC-C &	{Ka Ku X} &		19.1	\\
2013 Aug 22.2	 &	56526.2	&	488.2 &	19 &		13A-461	 	&	C	 &	X C L	 &	30.4	\\
2013 Aug 26.2	 &	56530.2	&	492.2 &	20 &		13A-461	 	&	C	 &	{Ka Ku X}	 &	19.1	\\

\hline
\end{tabular}\\
$^a$ Days after $t_0=$ 2012 April 21.0 = MJD 56038.0 \\
$^b$ Total time on source for all bands. \\
*P.I. C. C. Cheung.
\end{minipage}
\end{table*}

Since the discovery of $\gamma$-rays from novae by {the Fermi Gamma-ray Space Telescope}
({\it Fermi}) in 2010  \citep{Cheung10, Abdo10}, the production of early,
high-energy shocks in the ejecta of novae has become 
an increasingly important topic in the field of nova studies.  
These shocks can be caused by collisions with pre-existing circumstellar
material  {{\citep[eg., {V407~Cyg, V745~Sco, and RS~Oph};][]{Bode06, Das06, 
  Sokoloski06, Abdo10, Munari11, Chomiuk12, Nelson12, Banerjee14, Orio15}}}, or
interactions between multiple flows from the same eruption \citep[eg.,
  V959~Mon, T~Pyx,  V382 Vel;][]{Mukai01, Chomiuk14, Nelson14,
  Chomiuk14b}. Radio monitoring of novae, especially when combined with
observations at other frequencies, provides a powerful tool for
understanding the evolution of nova eruptions and examining shocks
within the ejecta.  High radio brightness temperatures can reveal
either non-thermal emission or shock-heated plasma,  as in the case of
V1723 Aql \citep{Krauss11, Weston13, Weston16}. Spatially resolved images of 
radio synchrotron emission in several novae have been used to trace the location of
relativistic particles that have been accelerated in shocks
\citep[eg., RS~Oph,
  V959~Mon;][]{OBrien06,Sokoloski08,Rupen08,Chomiuk14}. We show in
this paper that the 2012 nova V5589~Sgr is 
{one of the most extreme cases} known of a nova that is not
embedded in the wind of a red-giant companion having shock-powered
radio emission.  

V5589~Sgr (Nova Sgr 2012, PNV J17452791-2305213) was discovered to be
in outburst on 2012 April 21.0 ({MJD 56038.0}; all calendar dates 
are UT) by {\citet{CBET3089Korotkiy}}. 
For the rest of the paper, we take the
peak of the optical light curve to be 
MJD 56039.45
(which is roughly consistent with the peak in
HI1-B light curve from the Solar Terrestrial Relations Observatory, \citealt{CBET3089Korotkiy, Stereo2012};  {Eyres, in preparation}), and using data from
the Stony Brook / Small \& Moderate Aperture Research Telescope System (SMARTS)
Atlas of (mostly) Southern Novae \footnote{\url{http://www.astro.sunysb.edu/fwalter/SMARTS/NovaAtlas/}}
 \citep{Walter12} and the American Association of Variable Star Observers
(AAVSO), the time for the optical light to decrease by two magnitudes
was $t_2 = 6.2 \pm0.8$~days, and the time to decrease by three
magnitude was $t_3 = 12.8 \pm 1.5$ days. We take the start of the
eruption, $t_0$, to be 2012 April
21.0 (MJD  56038.0), near the beginning of the optical
brightening.

\begin{figure}
\centering 
\begin{minipage}{\columnwidth}
\caption{H$\alpha$ line profiles from selected SMARTS optical spectra of V5589~Sgr, colourized to show days after $t_0$=~MJD~56038.0  \citep{Walter12}. The spectra are normalized to the 6300-6400 \AA ~continuum, and are minimally smoothed with a Fourier filter to minimize high frequency noise. The rapidly evolving line profiles reveal that the ejecta likely had a complex flow structure.}
\includegraphics[width=67mm, angle=270]{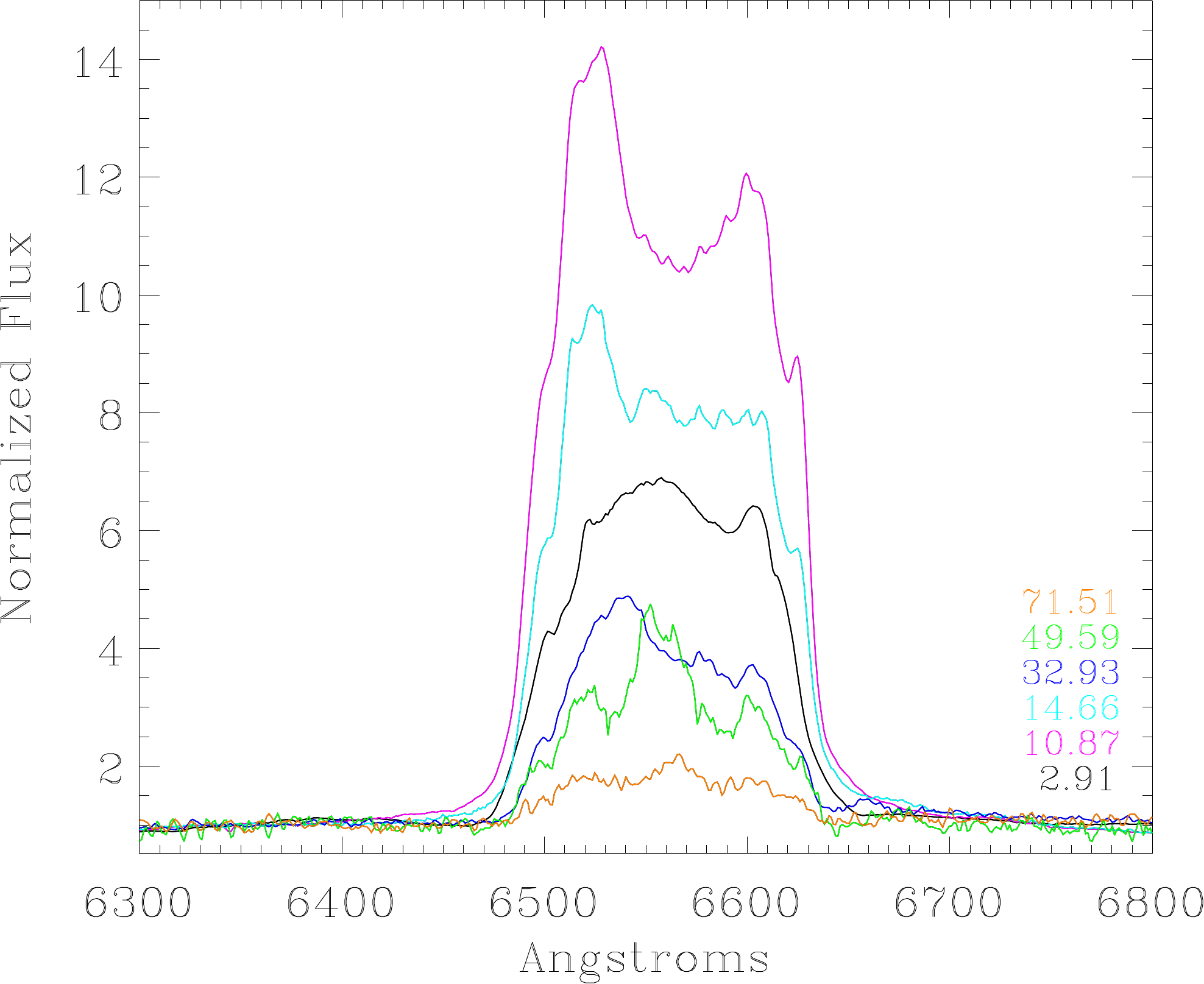}
\label{fig:OpticalSpectra}
\end{minipage}
\end{figure}

V5589~Sgr was classified as a fast `hybrid' nova, with
Fe II emission on day 3 and He/N features by day 8 \citep{ATEL4094,
  Walter12, Williams12}. 
The rapidly evolving, complex profiles
of the emission lines in the optical spectra from SMARTS
 reveal that the
ejecta were most likely aspherical and that they consisted of multiple
flows. The half width at zero intensity (HWZI) of the dominant
component in the H$\alpha$ and H$\beta$ emission lines from the SMARTS
spectra \citep{Walter12} during the first few days of the eruption was
${\rm v}_{\rm HWZI} = 4000{\rm \; km \; s}^{-1}$ 
(see Figure~\ref{fig:OpticalSpectra}).  
We refer to the outflow that
generated this emission component as the {\em fast flow}. 

 \citet{Mroz15} determined an orbital period of P = 1.59230(5) days using  {data from}
 the OGLE survey, and suggest that the system likely
contains a subgiant secondary and a massive white
dwarf.  This is a relatively unusual part of nova-progenitor parameter
space, with only a few such systems known \citep{Darnley12}. \citet{Mroz15}
additionally observed eclipses in the quiescent 
optical light curve, indicating that the inclination is close to
90$^\circ$ \citep{Mroz15}.  
The early line profiles of V5589~Sgr have some similarities to 
the recurrent nova U~Sco -- another eclipsing system which has a period of $\sim$1.2~days, 
and may contain an early K subgiant secondary \citep[see, e.g.,][]{Anupama13}.

On MJD 56038.9, \citet{ATEL4061} spent 1500~s on source with {\it
  Swift} / X-ray Telescope (XRT), finding no detectable X-rays and
placing an upper limit on 
the 0.3-10~keV flux of $<$0.02  
XRT counts per second\footnote{Their ATel actually quoted an upper
  limit of ``$>$'' 0.02 XRT counts per second, but we assume that the
  ``$>$'' sign was a typo.}; in UV, they found the source to have a magnitude of
$M2=13.90 \pm 0.05$. V5589~Sgr is not a known $Fermi$/LAT $\gamma$-ray
source: it is not in the 3FGL catalogue \citep{Acero15}, nor was it
detected as a transient in a blind search \citep{Cheung13}.    

In this paper, we describe how radio observations with the Karl
G. Jansky Very Large Array (VLA), X-ray observations with $Swift$/XRT,
and optical spectra from {SMARTS and} the Tillinghast Reflector Echelle
Spectrograph (TRES) at the Fred L. Whipple Observatory suggest that
the radio emission from V5589~Sgr consisted of two components: weak
thermal emission from a rapidly expanding photoionized plasma, and a
strong non-thermal, shock-driven flare. We describe the observations
in \S~2 and our results, along with a determination of the distance,
in \S~3. In \S~4, we describe the evidence that the radio flare was due
to non-thermal emission from particles accelerated in shocks, and
discuss the relationship between V5589~Sgr and $\gamma$-ray bright
novae such as V959~Mon. Finally, in \S~5 we summerise our conclusions.

\section{Observations}
\indent

\begin{sidewaystable}
\vspace*{8.6cm}
\caption{X-ray properties of V5589~Sgr. All uncertainties are 90 per cent confidence intervals.}
 \label{tab:Xrays}
\small
  \begin{tabular}{@{}llcccccccccccc@{}}
\hline
Observation date 	& MJD  		&  Day 	& Obs ID 			& Count rate 	 		& kT	\textsuperscript{\textdagger} & kT error & kT error &	 EM\textsuperscript{\textdagger \textdagger} 	 			& EM error 				& EM error 			& N(H)\textsuperscript{\textdagger} 	& N(H) error & N(H) error			 \\
				& 	  		&  	 	& 				&  {\it (counts per s) }			& {\it(keV)} & {\it (-)}	& {\it (+)}	 &{\it (cm$^{-3}$) }		& {\it (-)}					& {\it (+)}					&	{\it ($\times 10^{22}$cm$^{-2}$)} & (-) & (+)	 \\ \hline
2012 Apr 21.9		&  56038.91 	& 0.91 	& 00032442001 	& $<$0.01 			& ...		& ...		& ...		& ...					& ...						& ...						&	 ...	&	 ...	&	 ...			 \\
2012 Apr 25.7		&  56042.67 	& 4.7 	& 00032443001 	& $<$0.0021 			& ...		& ...		& ...		& ...					& ...						& ...						&	 ...	&	 ...	&	 ...			 \\
2012 Apr 26.1 		& 56043.13 	& 5.1 	& 00032443002 	& $<$0.0078 			& ...		& ...		& ...		& ...					& ...						& ...						&	 ...	&	 ...	&	 ...			 \\
2012 May 10.5		& 56057.50  	& 19.5  	& 0032443003 		& 0.045 $\pm$ 0.003 	&  $>$32.7 & ...		& ... 		&	3.62$\times10^{56}$	&	-3.62$\times10^{55}$	&	+5.43$\times10^{55}$ 	& 	0.43 & -0.30 & 0.60			\\
2012 May 17.7 		& 56064.66  	&  26.7 	& 00032442002 	& 0.083 $\pm$ 0.010 	&  $>$5.2  & ...		& ...		&	5.43$\times10^{56}$	&	-1.27$\times10^{56}$	&	+1.99$\times10^{56}$ 	&  	$<$0.68 &	 ... & ...			\\
2012 May 25.6		& 56072.60  	& 34.6  	&  00032442004 	& 0.091 $\pm$ 0.011 	& $>$3.5  	& ...		& ...		&	6.15$\times10^{56}$	&	-2.71$\times10^{56}$	&	+1.63$\times10^{56}$	&  	0.52 & -0.09 & 1.04				\\
2012 May 28.6	 	& 56075.55 	& 37.6 	& 00032442005 	& 0.055 $\pm$ 0.008 	& $>$3.1  	& ... 		& ... 		&	2.53$\times10^{56}$	&	-7.23$\times10^{55}$	&	+7.23$\times10^{55}$	& 	$<$ 0.18& ...  & ...			\\
2012 May 31.6	 	& 56078.62  	&  40.6 	& 00032442006 	& 0.074 $\pm$ 0.009 	& 4.3 	& -1.9 	& +11.0 	&	4.16$\times10^{56}$	&	-1.63$\times10^{56}$	&	+1.27$\times10^{56}$	&  	0.32 & -0.06 & 0.67			\\
2012 Jun 6.9		&  56084.91 	& 46.9  	& 00032442007 	& 0.077 $\pm$	0.015	& 1.9 	& -0.9	& +9.2 	&	3.26$\times10^{56}$	&	-1.63$\times10^{56}$	&	+1.09$\times10^{56}$	& 	$<$1.19 & ...  &...			\\
2012 Jun 10.1	 	& 56088.10 	& 50.1  	& 00032442008/9 	& 0.038 $\pm$ 0.003 	&  1.3 	&  -0.1	& +0.1	&	2.53$\times10^{56}$	&	-5.43$\times10^{55}$	&	+5.43$\times10^{55}$	& 	0.36 & -0.23  &0.55			\\
2012 Jun 12.8	 	&  56090.84 	& 52.8  	& 00032442010 	&  0.061 $\pm$ 0.008	& 3.5 	& -1.4 	& +4.4 	& 	2.53$\times10^{56}$	&	-7.23$\times10^{55}$	&	+3.63$\times10^{55}$	& 	9.5$\times 10^{-2}$ & -0.009  &0.22			\\
2012 Jun 23.5		& 56102.49  	&  64.5  	& 00032442011 	& 0.061 $\pm$ 0.010 	& 0.14 	& -0.05	& +0.14 	&   $<$$1.85\times10^{60}$	&	... 					&	...				& 	0.51 & -0.31 & 0.75	 		\\
2012 Jun 28.3 		& 56106.34 	& 68.3 	& 00032442012 	& 0.071 $\pm$ 0.009 	& 1.37 	& -0.38 	& $+0.62$ &	 1.45$\times10^{56}$ &	-9.04$\times10^{55}$	&	+7.23$\times10^{55}$	& 	0.42 & -0.39 & 0.64			\\
2012 Jul 8.1 		& 56116.10  	& 78.1  	& 00032442013 	& 0.043 $\pm$ 0.008 	& \multicolumn{4}{r|}{too few counts for spectroscopy}													&	 ...	&	 ...	&	 ...	&	 ...	&	 ...			\\
2012 Jul 10.8 		& 56118.84 	& 80.8  	& 00032442014 	& 0.065 $\pm$ 0.010 	& \multicolumn{4}{r|}{too few counts for spectroscopy}													&	 ...	&	 ...	&	 ...	&	 ...	&	 ...			\\
2012 Jul 14.8 		& 56122.80 	& 84.8  	& 00032442015 	& 0.034 $\pm$ 0.013 	&  \multicolumn{4}{r|}{too few counts for spectroscopy}													&	 ...	&	 ...	&	 ...	&	 ...	&	 ...			\\
2012 Aug 7.3  		& 56146.25 	&108.3	& 00032442017	& $<$0.011 			& ...		& ...		& ...		& ...					& ...						& ...						&	 ...	&	 ...	&	 ...			 \\ \hline
\end{tabular}\\
{\textdagger Values of kT and the column density, N(H), are from model fits using the {\tt tbabs*apec} model in Xspec.}\\
{\textdagger \textdagger  We used a distance of 4 kpc in calculations of the emission measure, EM.}
\end{sidewaystable}

The VLA observed V5589~Sgr beginning three days after initial
discovery through the 16 months following with the WIDAR correlator at
frequencies between 1 and 37~GHz. The observations used 8-bit
samplers with band widths of 2048~MHz in two tunable 1048~MHz
sub-bands at all frequencies except {\it L}-band, which had a
bandwidth of 1024 MHz split into two sub-bands. VLA
observations of V5589~Sgr were alternated with a nearby phase
reference calibrator (J1751-2524 or J1755-2232).  Each observation
additionally included the observation of a standard flux calibrator
(3C~48 or 3C~286). VLA observations were reduced either using
Astronomical Image Processing System (AIPS) and following the standard
procedures, or with the Common Astronomical Processing System
(CASA) v4.2.2 \citep{McMullin07}, using the VLA calibration pipeline
v1.3.1. For imaging, we used the CLEAN algorithm \citep{Hogbom74} with
Briggs robust weighting of 0.5. Based on the distance and size of the source, 
as well as the timing of our observations, we did not expect to resolve the 
ejecta from V5589~Sgr, and this proved to be the case. For all detections, we determined flux
densities by fitting the source with a gaussian using the {\it imfit}
task in CASA or the {\it JMFIT} task in AIPS. For non-detections, 
we either used our usual imaging routines to determine the RMS and flux density
at the target location, or we used {\it difmap} \citep{Shepherd97} to
find the flux density at the target location and the off-source
RMS. We estimated uncertainties in flux density by adding the error
from the gaussian fits in quadrature with estimated systematic errors of 5 per cent
below 19~GHz and 10 per cent above, with 1$\sigma$ error bars from error
propagation thereafter.  For all non-detections we quote upper limits
as 2$\times$RMS.  Table~\ref{tab:Observations}
lists the VLA observations. To determine the spectral index $\alpha$
(where $S_\nu \propto \nu^\alpha$ and $S_\nu$ is the flux density at
frequency $\nu$) of the emission at each epoch, we 
used IDL's {\it linfit} function to fit the data {with} a power-law {model} in log
space.

Additionally, we obtained two observations at 225~GHz on the
Submillimeter Array (SMA) under program 2012A-S016. Observations with
the SMA were both obtained in compact configuration, with 60 minutes
and 21 baselines on MJD 56069.0 (day 31.0) and 208 minutes and 15
baselines on MJD 56117.9 (day 79.9).  These data were reduced using
standard routines in IDL and Miriad.

We also monitored the V5589~Sgr outburst using the
XRT instrument onboard the {\it Swift} satellite, which resulted in a
series of 17 observations carried out over 15 weeks.  An additional
follow-up observation was obtained on 2014 April 21.  All observations
were carried out in the XRT's photon counting (PC) mode, and had
durations ranging from 300 to 6300~s (see Table~\ref{tab:Xrays}).  We
extracted spectra following the usual reduction prescription for XRT
data -- counts were extracted from a circular region with a radius of 5\arcsec centred on the source using Select v.2.4c.  Background counts were extracted from a larger circular region (radius 155\arcsec) away from the source.  Ancillary response files were produced using the {\it xrtmkarf} tool, correcting for the profile of the psf. Finally, we used the latest response matrix file from the Swift calibration database (swxpc0to12s6\_20110101v014.~rmf). In observations where V5589~Sgr was not detected, we utilised the Bayesian method outlined in \citet{Kraft91} to calculate upper limits on the count rate. 

We obtained optical spectra of V5589~Sgr using the Tillinghast
Reflector Echelle Spectrograph (TRES) on the 60-inch reflector at the
Fred L. Whipple Observatory on MJD 56046.0 (day 8)
and MJD 56057.0 (day 19). We reduced the data using the standard
reduction procedure \citep{Mink11}.

\section{Results}

\begin{table*}
  \caption{Radio flux densities for V5589~Sgr from the VLA. Quoted
    uncertainties consist of RMS and systematic errors added in
    quadrature.  For non-detections, we quote upper limits of $2\times
    RMS$. Day number reflects the number of days after $t_0 =$~MJD
    56038.0.} 
 \begin{minipage}{230mm}
  \label{tab:Flux}
  \small
  \begin{tabular}{@{}ccccccccccc@{}}
  \hline

 {\bf Day}  & \multicolumn{10}{c}{\bf Observed flux density (mJy)}\\
& \multicolumn{1}{r|}{}{\it 1.4 GHz} & \multicolumn{1}{r|}{\it 1.8 GHz} & \multicolumn{1}{r|}{\it 4.7 GHz } & \multicolumn{1}{r|}{\it 7.6 GHz} & \multicolumn{1}{r|}{\it 8.5 GHz } &\multicolumn{1}{r|}{\it 11.4 GHz } & \multicolumn{1}{r|}{\it 13.3 GHz} & \multicolumn{1}{r|}{\it 17.5 GHz} & \multicolumn{1}{r|}{\it 27.5 GHz} & \multicolumn{1}{r|}{\it 36.5 GHz}  \\ 
& \multicolumn{1}{r|}{}{\it  (L band)} & \multicolumn{1}{r|}{\it (L band)} & \multicolumn{1}{r|}{\it (C band)} & \multicolumn{1}{r|}{\it(C band)} & \multicolumn{1}{r|}{\it  (X band)} &\multicolumn{1}{r|}{\it  (X band)} & \multicolumn{1}{r|}{\it  (Ku band)} & \multicolumn{1}{r|}{\it (Ku band)} & \multicolumn{1}{r|}{\it  (Ka band)} & \multicolumn{1}{r|}{\it  (Ka band)} \\
\hline
	2.4 		& 	...	&	...	&	...	&	...	&	...	&	...	&	...	&	...	&	...	&	$<$0.05$^{a}$		\\ 
	46.4 		&	...	&	...	&1.09 $\pm$ 0.06$^{b}$ &	1.38 $\pm$ 0.07$^{c}$&	...	&	...	&	...	&	...	&	...	&	...		 	\\ 
	54.2 		&	...	&	...	&	...	&	...	&	...	&	...	&	3.52$\pm$0.18 &	3.98$\pm$0.20	&5.73$\pm$0.58	&5.96$\pm$0.61	 	\\ 
	62.3 		&	3.09$\pm$0.16$^{d}$&	3.20$\pm$0.16	$^{e}$&3.42$\pm$0.17 $^{b}$&	3.63$\pm$0.18$^{c}$	&3.68$\pm$0.19 &	3.78$\pm$0.20 &	...	&	...	&	...	&	...		\\ 
	81.1 		&	1.68$\pm$0.15	&1.39$\pm$0.10	&1.91$\pm$0.10	&1.92$\pm$0.10	&...	&	...	&	...	&	...	&	...	&	...	 	\\ 
	90.3 		&	...	&	...	&	...	&	...	&	...	&	...	&	...	&	1.37$\pm$0.11&	1.44$\pm$0.20&	1.14$\pm$0.22	 	\\ 
	110.0 	&	0.34$\pm$0.07	& 0.56$\pm$0.08 & 0.52$\pm$0.04 & 0.53	$\pm$0.04 & 0.56$\pm$0.04 & 0.57$\pm$0.07&	...	&	...	&	...	&	...		\\ 
	112.2 	&	...	&	...	&	...	&	...	&	...	&	...	&	0.50$\pm$0.05	& 0.38$\pm$0.05&	0.47$\pm$0.09&	0.39$\pm$0.09 	\\ 
	113.3 	&	$<$0.18&	0.50$\pm$0.09&	0.50$\pm$0.04	& 0.47$\pm$0.03 &	0.44$\pm$0.03	& 0.42$\pm$0.04 &	...	&	...	&	...&	...	 	\\ 
	129.0 	&	0.24$\pm$0.05&	0.43$\pm$0.05	& ... &	...	& ...	& ... &	...	&	...	&	...	&	...		 	\\ 
	129.0 	&	$<$0.15 &	... &0.35$\pm$0.03&	0.29$\pm$0.02	&0.29$\pm$0.02	&0.37$\pm$0.04&	...	&	...	&	...	&	...		 	\\ 
	129.2 	&	$<$0.17 &	0.35$\pm$0.09	&0.28$\pm$0.04&	0.29$\pm$0.03	& ...	& ... &	...	&	...	&	...	&	...		 	\\ 
	136.2 	&	...	&	...	&	...	&	...	&	...	& ...	 &	0.19$\pm$0.02 &	0.11$\pm$0.03	 & 0.11$\pm$0.03 &	0.37$\pm$0.14 	\\ 
	140.0 	&	0.23$\pm$0.05 &	0.20$\pm$0.04&	...	&	...	&	...	&	...	&	...	&	...	&	...	&	...		\\ 
	193.0 	&	...	&	...	&	...	&	...	&	...	&	...	&	$<$0.05	&	$<$0.05	&	$<$0.09	&	$<$0.10			\\ 
	232.8 	&	$<$0.17	&	$<$0.17	& $<$0.04	&	$<$0.03	&	$<$0.04	&	$<$0.06	&...	&	...	&	...	&	...		 	\\ 
	248.1 	&	...	&	...	&	...	&	...	&	...	&	...	&	$<$0.04	&$<$0.05	&	$<$0.10&	$<$0.14			\\ 
 	259.7 	&	$<$0.20 &		$<$0.14	&	$<$0.07	&	$<$0.05	&$<$0.04	&	$<$0.05	&	...&	...	&	...&	...		\\ 
	309.7 	&	$<$1.01	&	$<$0.74	&	$<$0.23	&$<$0.32	&$<$0.40	&	$<$0.60	&...	&...	&	...&	...	\\ 
	 320.5 	&	...	&	...	&	...	&	...	&	...	&	...	&	$<$0.05	&	$<$0.05	&	$<$0.10	&	$<$0.12			\\ 
	321.6 	&	...	&	...	&	...	&	...	& ...	&	...	&	$<$0.05	&	$<$0.06	&	$<$0.13	&	$<$0.17			\\ 
	403.4 	&	$<$0.79	&	$<$0.65	&$<$0.07&$<$0.04&	$<$0.04	&	$<$0.04	&	...	&	...	&	...	&	...		\\ 
	408.2 	&	...	&...	&	...	&	...	&	...	&	...	&	$<$0.05	&	$<$0.06	&	$<$0.15	&	$<$0.19			\\ 
	488.2 	&	$<$0.55 &	$<$0.34	&$<$0.06	&	$<$0.04	&	$<$0.04	&$<$0.05	&	...	&	...	&	...	&	...		 	\\ 
	492.2 	&	...	&	...& ...	&	...	&...	&	...	&	$<$0.06	&	$<$0.11	&	$<$0.02	&	$<$0.22		\\ 
\hline\\
\end{tabular} \\

$^{a}$ Observed at 33~GHz on this date  \\
$^{b}$ Observed at 5.0~GHz on this date\\
$^{c}$ Observed at 6.8~GHz on this date \\
$^{d}$ Observed at 2.5~GHz ({\it S}-band) on this date \\ 
$^{e}$ Observed at 3.5~GHz ({\it S}-band) on this date \\
\end{minipage}
\end{table*}

\subsection{Optical emission and estimate of distance}

\begin{figure}
\centering 
\begin{minipage}{\columnwidth}
\caption{{Diffuse Interstellar Band (DIB) features from the TRES spectra of V5589~Sgr, which furnished constraints on reddening value and distance.}}
\includegraphics[width=\columnwidth]{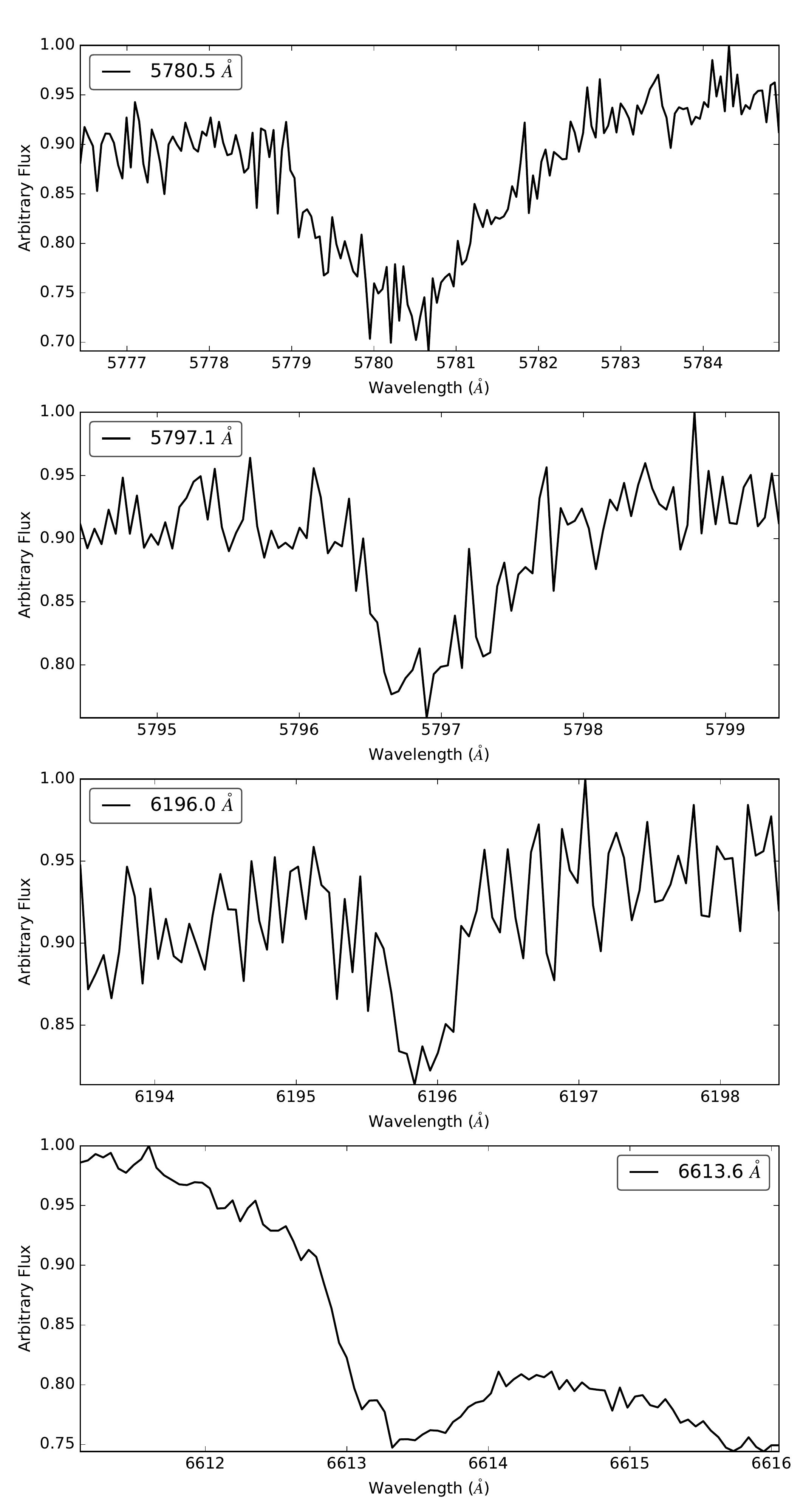}
\label{fig:DIB}
\end{minipage}
\end{figure}

\begin{figure}
\centering 
\begin{minipage}{\columnwidth}
\caption{Relationship between the distance modulus and the peak
  absolute visual
  magnitude ($M_V$) of V5589~Sgr, using reddening derived 
from DIB features {shown in Figure}~\ref{fig:DIB} \citep{Finzell15, Green15}.  The horizontal dashed lines show
the expected range of $M_V$ of between -9.0 and -7.0 mag, and the
vertical lines show the 
corresponding distance moduli of between 12.5 and 14.5.}
\includegraphics[width=\columnwidth]{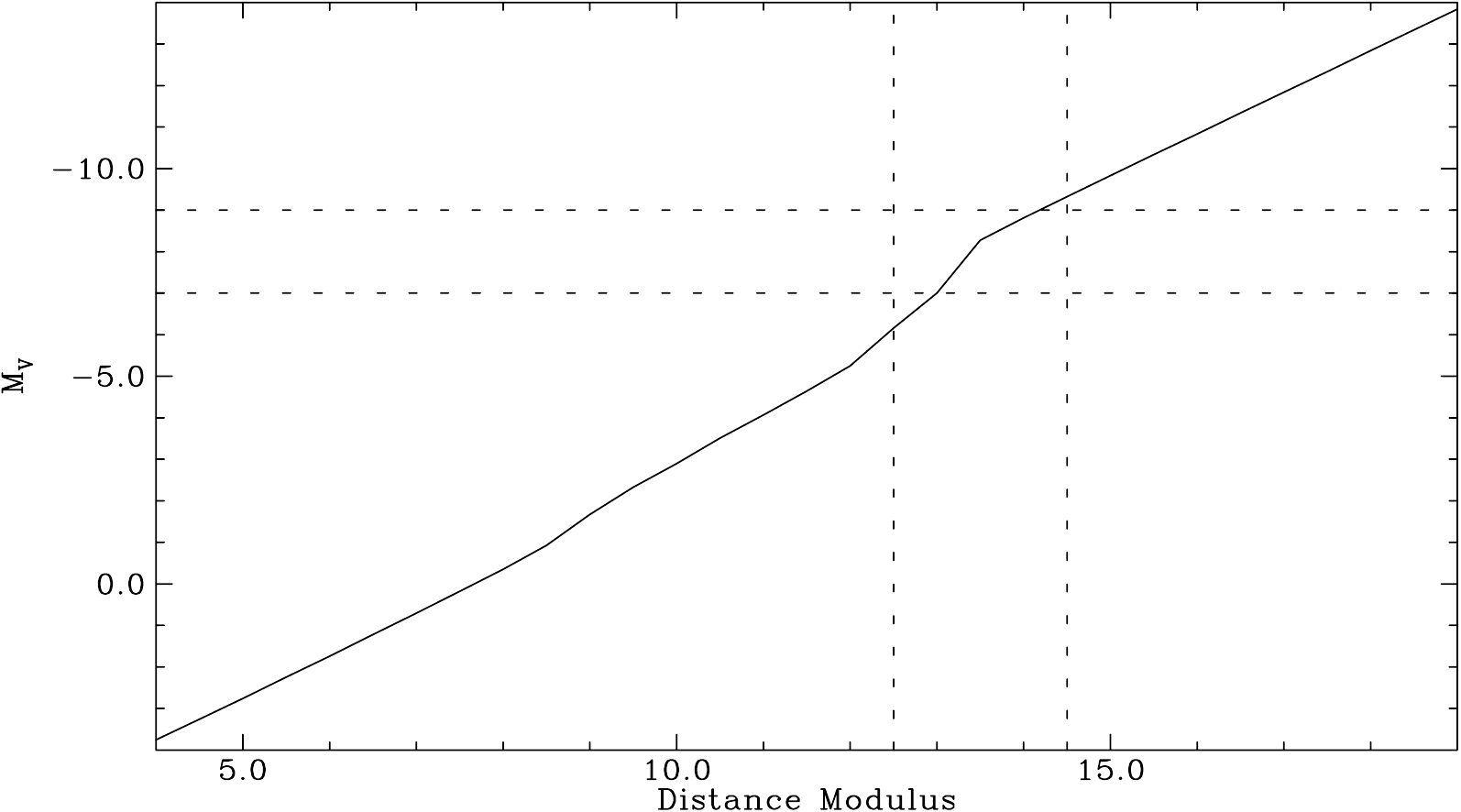}
\label{fig:Distance}
\end{minipage}
\end{figure}

\indent

To obtain a first estimate of the distance to V5589~Sgr, we used optical
spectroscopy guided by the typical peak optical brightness of
novae. From the TRES spectra, we derived a reddening value by
measuring the equivalent width of diffuse interstellar band (DIB)
features and using the calibration technique
of~\citet{Friedman11}, {a method which is independent of any assumptions about the properties of the nova}. 
Using four DIB features (5780~\AA, 5796~\AA,
6195~\AA, and 6613~\AA, see Figure~\ref{fig:DIB}), we obtained a reddening of $E(B-V) = 0.8 \pm
0.19$ \citep[see][for details of this technique; we were unable to use
  the Na  {\small I} D  lines at $5889.9~\AA$  and $5895.9~\AA$ due to
  saturation, or the K {\small I} line at 7698.9 \AA, which fell into
  a chip gap in the CCD]{Finzell15}. This reddening value, in
conjunction with the 3D reddening map of~\citet{Green15}, suggests a
distance of 3.6$^{+1.0}_{-2.4}$~kpc.  However, taking into account
the uncertainties on the reddening of V5589~Sgr and in the reddening
map, the most likely 
distance
modulus is 12.9 $\pm$ 1.4. This value corresponds to a distance of
$\sim  3.8$~kpc.

The peak optical
brightness of V5589~Sgr is consistent with a distance 
of around 4~kpc. The distance and the 
extinction, $A_V$, are strongly coupled. Taking the peak apparent V magnitude
to be 8.8 \citep{CBET3089Korotkiy}, extinction given by $A_V=3.1
\times E(B-V)$, and 
therefore peak absolute {magnitude} of $M_V=8.8-A_V-DM$, we find
$M_V$ as a function of distance modulus 
(see Figure~\ref{fig:Distance}). 
Despite a limiting magnitude of $M_R\sim -4.4$, the faintest nova in
the \citet{Cao12} Palomar Transient factory list of M31 novae had an $M_V$
of $-6.5$, with the vast majority between $M_V = -7.0$ and $M_V =
-9.0$. Therefore, 
even if we assume that V5589~Sgr was a relatively faint nova with
$M_V=-7.0$, we can eliminate the lowest part of the
distance modulus range that is formally allowed by the 3D reddening
map. Taking the distance constraints from both reddening 
and peak optical brightness into consideration, it is unlikely that
V5589~Sgr is closer than 3.2~kpc (a distance modulus of 12.5), at which
$M_V=-6.2$. We therefore take the distance of the source to be
approximately 4~kpc (or within the range 3.2 to 4.6~kpc). 

%

\subsection{Radio emission}
\begin{figure}
\centering 
\begin{minipage}{\columnwidth}
\caption{Radio flux densities for V5589~Sgr taken with the VLA. Error bars are as reported in Table~\ref{tab:Flux}, but may be too small to be visible. Upper limits are $2\times RMS$ on non-detections. {These data are also shown in Figure}~{{\ref{fig:Flux}}}.}
\includegraphics[width=\columnwidth]{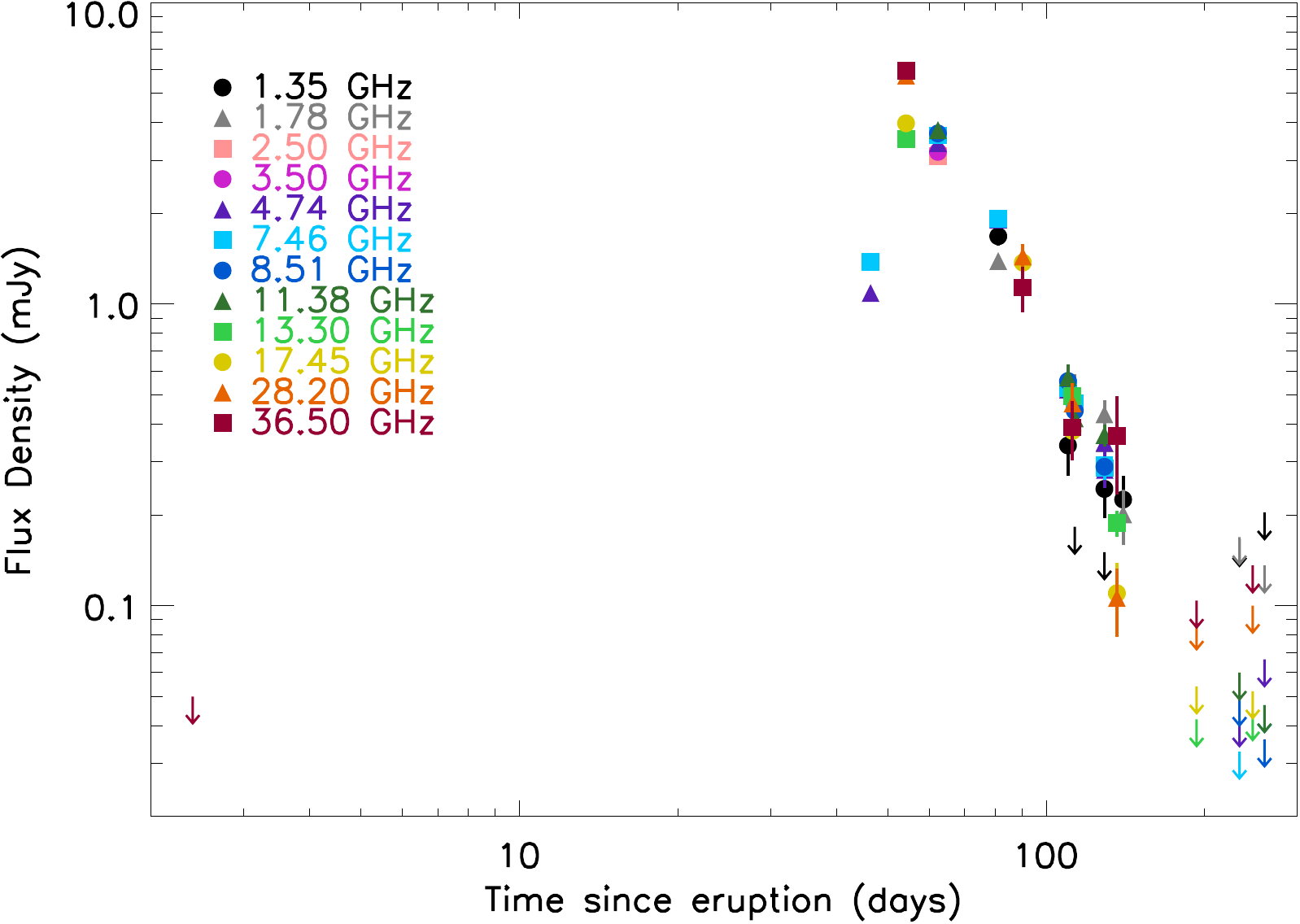}
\label{fig:FluxOnly}
\end{minipage}
\end{figure}

\indent

{A month} after an early radio non-detection on day~2.4, V5589~Sgr brightened
exceptionally quickly at low frequencies to a peak approximately 2~months after the start of the eruption 
(see Figure~\ref{fig:FluxOnly}).  
Although we were not able to obtain low-frequency ({\it L} and {\it C}-band)
observations between days 2.4 and 46.2, the
flux rose as $t^{3.9\pm 0.1}$ at 
5.0~GHz and as $t^{3.3\pm 0.1}$ at 6.8~GHz {between days 46.4 and 62.3} ({where t is the time from} $t_0$, again using MJD 56038.0 as $t_0$).  
The speed of the low-frequency radio brightening is noteworthy because the
strength of thermal emission from a source with constant temperature
is proportional to the size of the source on the sky; for a
freely expanding source, the size cannot increase faster than $t^2$.
The VLA did not obtain
observations at high frequencies during the rapid, low-frequency brightening. 
The flux densities reached global maxima in {\it 
  Ku}-band (13.3/17.5~GHz) and {\it Ka}-band (27.5/36.6~GHz) on day
54.2.  {\it L}-band (1.2/1.8~GHz), {\it C}-band
(4.7/7.6~GHz), and {\it X}-band (8.5/11.2~GHz) experienced their peak
flux densities on day 62.3.  Observations with the SMA at 225~GHz
revealed a flux density of 13.80$\pm$2.43 mJy on day 31.0 and an upper
limit of 0.14 mJy on day 79.9.  After these epochs, the flux densities
fell monotonically at all frequencies (see Figure~\ref{fig:FluxOnly}
and Table~\ref{tab:Flux}). 
The flux fell as $t^{-3.2 \pm 0.2}$ at
1.2~GHz between days 62.3 and 140.0 and as $t^{-3.0 \pm 0.5}$ at
36.5~GHz between days 54.2 and 136.2. After day 140.0, 
all observations 
resulted in non-detections. 
  
\begin{figure*}
\centering 
\begin{minipage}{160 mm}
\caption{Spectra of V5589~Sgr between day 46.2 and 140.0 after ejection, with linear spectral fitting for spectral index $\alpha$ at each epoch and 1$\sigma$ error. }
\includegraphics[width=163 mm]{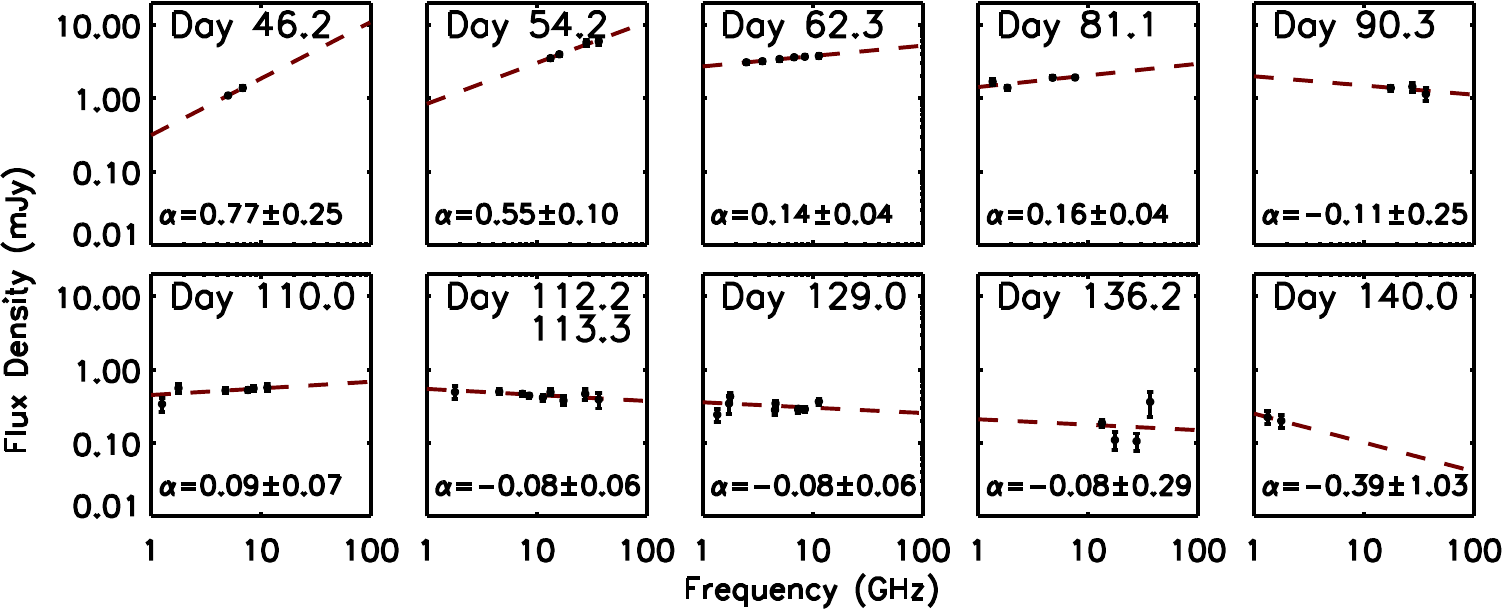}
\label{fig:Spectra} 
\end{minipage}
\end{figure*}

Because the low-frequency
{emission strengthened} so quickly, 
the radio spectrum 
flattened as the source brightened (see Figure~\ref{fig:Spectra}). 
The radio spectrum initially rose towards high frequencies with a
spectral index of $\alpha =0.77 \pm 0.25$ between 5.0 and
6.8~GHz on day 46.2.  On day 54.2, the spectrum had flattened to
$\alpha =0.55 \pm 
0.10$ between 13.0 and 36.5~GHz.  And by day 62.3, it had flattened further
to $\alpha =0.14 \pm 0.04$ between 2.5 and 11.4~GHz.  Thereafter, it
remained roughly flat or falling with frequency for all subsequent
observations (see Figure~\ref{fig:Spectra}).  
{A freely expanding isothermal remnant (as in the so-called} {\em Hubble-flow model}) {is expected to remain completely optically thick during the initial period of radio brightening, producing the characteristic steeply rising spectrum of a blackbody in the Rayleigh-Jeans limit} 
\citep[e.g.,][]{Seaquist77, Hjellming79, Bode08}.
 Indeed, radio emission
from novae during the early stages of expansion of the remnant often
has a spectral index $\alpha > 1$ \citep[e.g., 1.5 at low frequencies
  for V705~Cas, 1.6 for V339~Del;][]{Eyres00,ATEL5382}.
The spectral index of emission from such a freely expanding, thermal
remnant then flattens -- first at the highest and then at lower
frequencies -- to $\alpha = -0.1$ as the ejecta become optically thin
\citep{Bode08}.
The flattening of the radio spectrum from V5589~Sgr as the radio flux
rose is thus one of the first hints that the radio emission {was} not all
thermal. 


Moreover, the VLA observations uncovered possible signs of linear polarization,
as would be expected from non-thermal, synchrotron emission.
While we did not observe a polarization calibrator during the peak of the flux
density, there was a standard VLA polarization TPOL observation several days prior
to our observation at similar, though not identical, frequencies in {\it Ku}-band.
Using this dataset and our observation of 3C286, we were able to {perform} rough polarization calibrations for {\it Ku} and {\it Ka}-bands, albeit with large margins of uncertainty. On day 54.2,  we found upper limits to polarization in Ka-band of $<$0.116~mJy ($<2.0$ per cent) at 27.5~GHz, and $<$0.122~mJy ($<2.0$ per cent) at 36.5~GHz. In Ku-band, we found a possible { } $\sim$0.2~mJy (5 per cent) of linearly polarized flux. While this detection is of a similar order to our uncertainty at {\it Ku}-band if we include our systematic error factor, it could hint at the presence of synchrotron emission.




To investigate the surprising radio spectral evolution and fast rise, we
tested whether any of the radio emission could be described by a
`Hubble-flow' type  
ejection \citep[see, e.g.,][]{Seaquist77, Hjellming79, Weston16}  
{with velocity proportional to the distance from the source, $r$,} between
 \v$_{min}$ and \v$_{HWZI}$ and a $r^{-2}$ density profile. 
We take any Hubble-flow to have begun around $t_0$.
Although a linear extrapolation of the optical light curve backward in
time suggests that the thermonuclear runaway could have occurred a day
or two before our chosen $t_0$, the VLA observations support a start
of the fast, 4,000~km~s$^{-1}$ flow on or after $t_0$; 
if the fast flow started before $t_0$ and remained very optically
thick at all radio frequencies for at least a few days, the flux
density at 33~GHz would have been \\ \\ 
$S_{\nu}(t) > 0.07\; {\rm mJy} \; $
\begin{equation}
\times \left( \frac{T}{10^4~{\rm K}} \right) \left(
\frac{\nu}{33~{\rm GHz}} \right)^2 \left( \frac{{\v_{HWZI}}}{4000\,{\rm
    km\, s}^{-1}} \right)^2 \left( \frac{t}{2.4~{\rm 
    day}} \right)^2 \left( \frac{D}{4~{\rm kpc}} \right)^{-2},
\end{equation}
where $T$ is the temperature of the ejecta.  
The first VLA observation, however, produced a non-detection at
33~GHz, with a 3$\sigma$ upper limit of 0.08 mJy on day 2.4
(MJD~56040.4), suggesting that the fast flow did not start before $t_0$.
%
Because the 4,000~km~s$^{-1}$ flow {was} clearly evident in optical
spectra on day 2.0 \citep{CBET3089Buil} and day 2.4 \citep{Walter12},
it must therefore have started between $t_0$ (day 0) and day 2.  

\begin{figure*}
\centering 
\begin{minipage}{180 mm}
\caption{Radio flux densities { }and
  brightness temperatures for V5589~Sgr through day 300. 
{\bf Top:} {Flux density at high frequencies (7.45-255.3 GHz).} {\bf Middle:} {Flux density at low frequencies (1.35-4.74 GHz).} {\bf Low:} {Brightness temperature at all frequencies.}
  Although the radio emission  after day 100 is roughly consistent with expectations from a freely
  expanding $10^4$~K remnant, the high brightness temperatures and
  poor model fits at low frequencies before this time show that an additional source of radio emission was present before day 100. We argue in the text that this radio flare was due to synchrotron emission from particles accelerated in shocks. The model fits (dashed lines) are for a Hubble Flow model with $D=4$~kpc, \v$_{HWZI}=$ 4000 km s$^{-1}$, T$=1.2\times 10^{4}$ K, M$_{ej}= 2.6 \times 10^{-5}$ M$_{\odot}$,  $\zeta {\equiv}\frac{\v_{min}}{\v_{HWZI}}=0.84$, and ejection at $t_0=$ 56038.0 MJD. All observations after day 140.0 were non-detections.}

\centering
\includegraphics[width=110 mm]{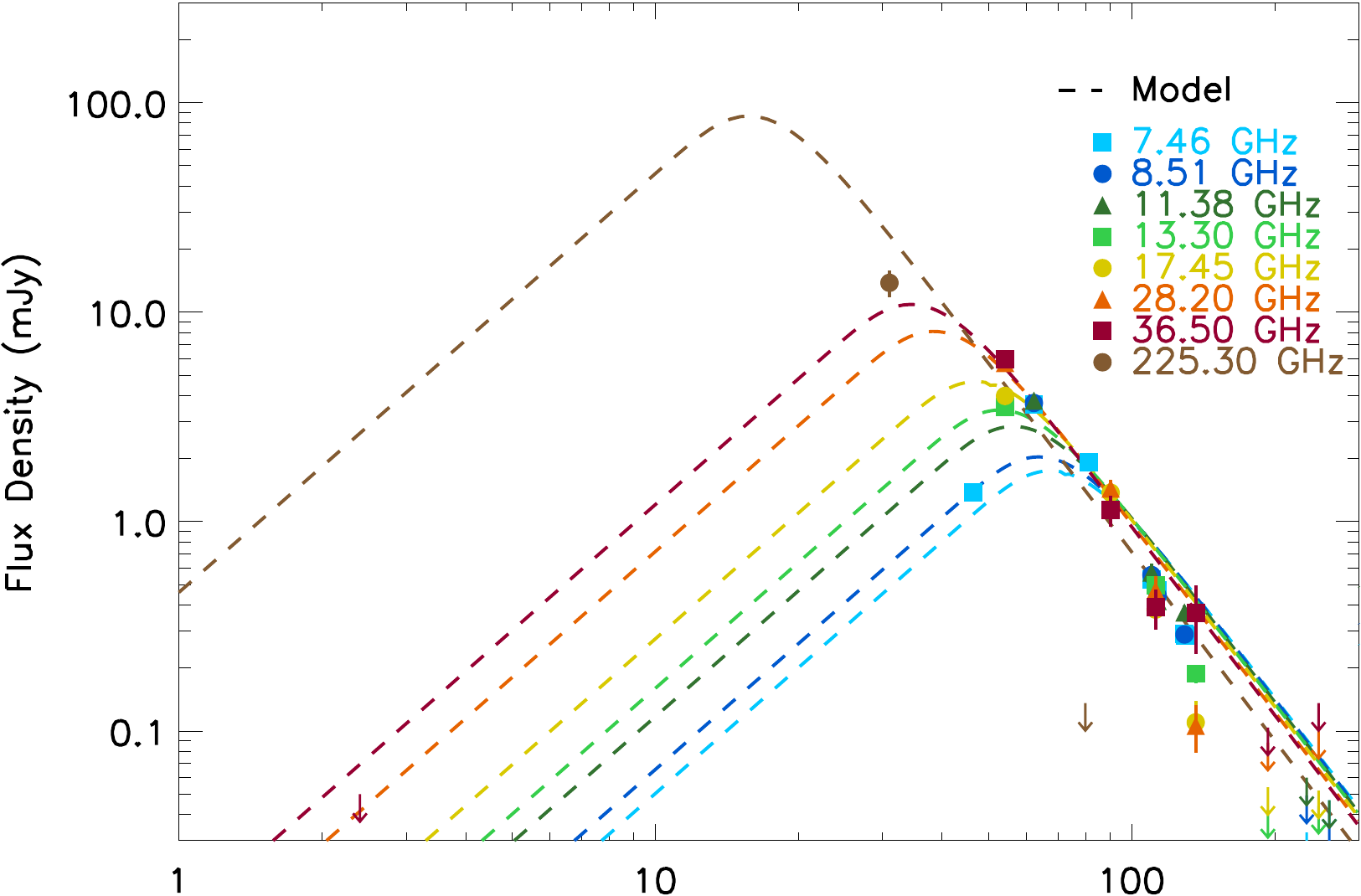}
\vspace*{-0.36cm}

\includegraphics[width=110 mm]{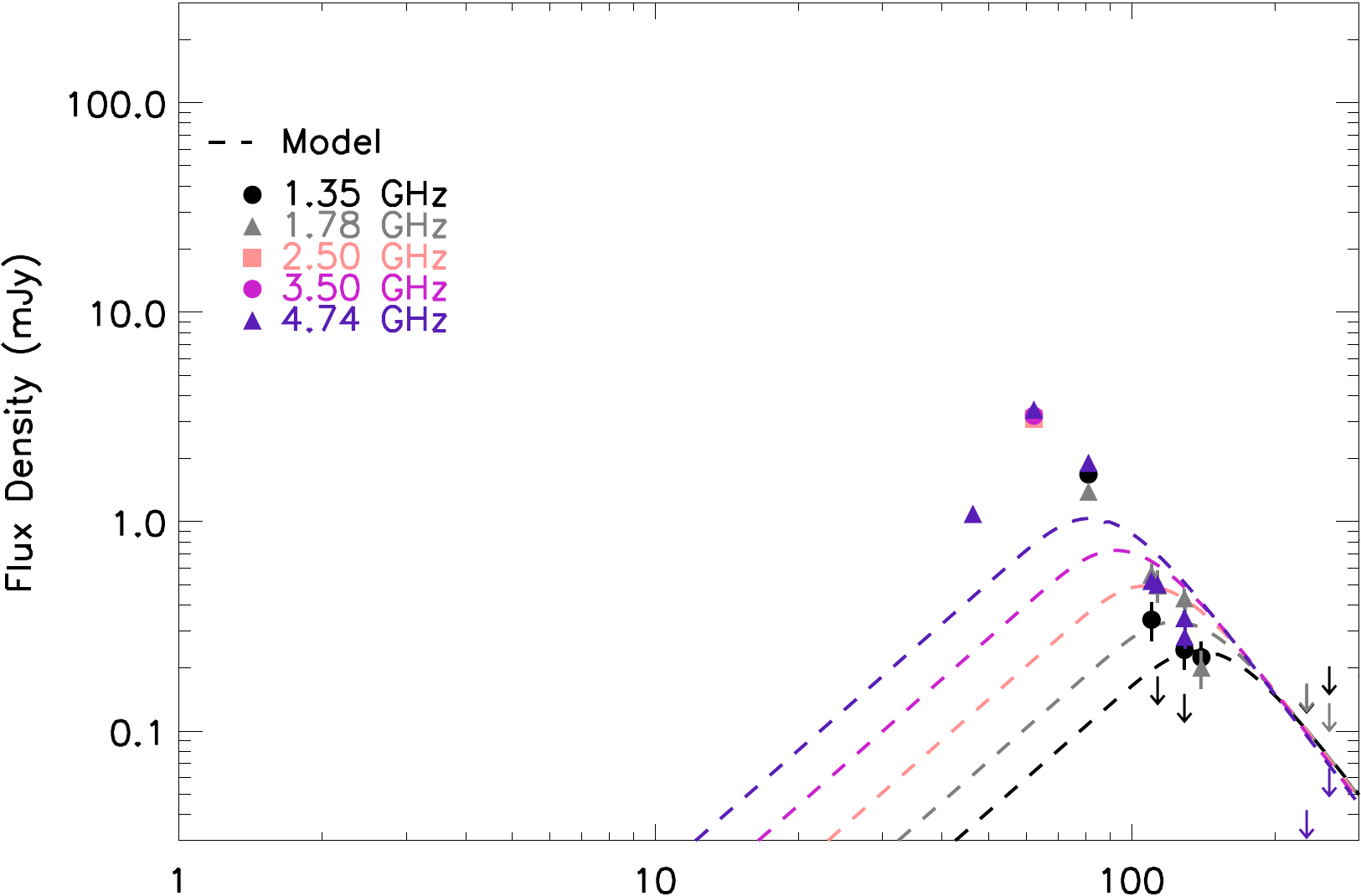}
\vspace*{-0.35cm}

\hspace*{0.2cm}
\includegraphics[width=107 mm]{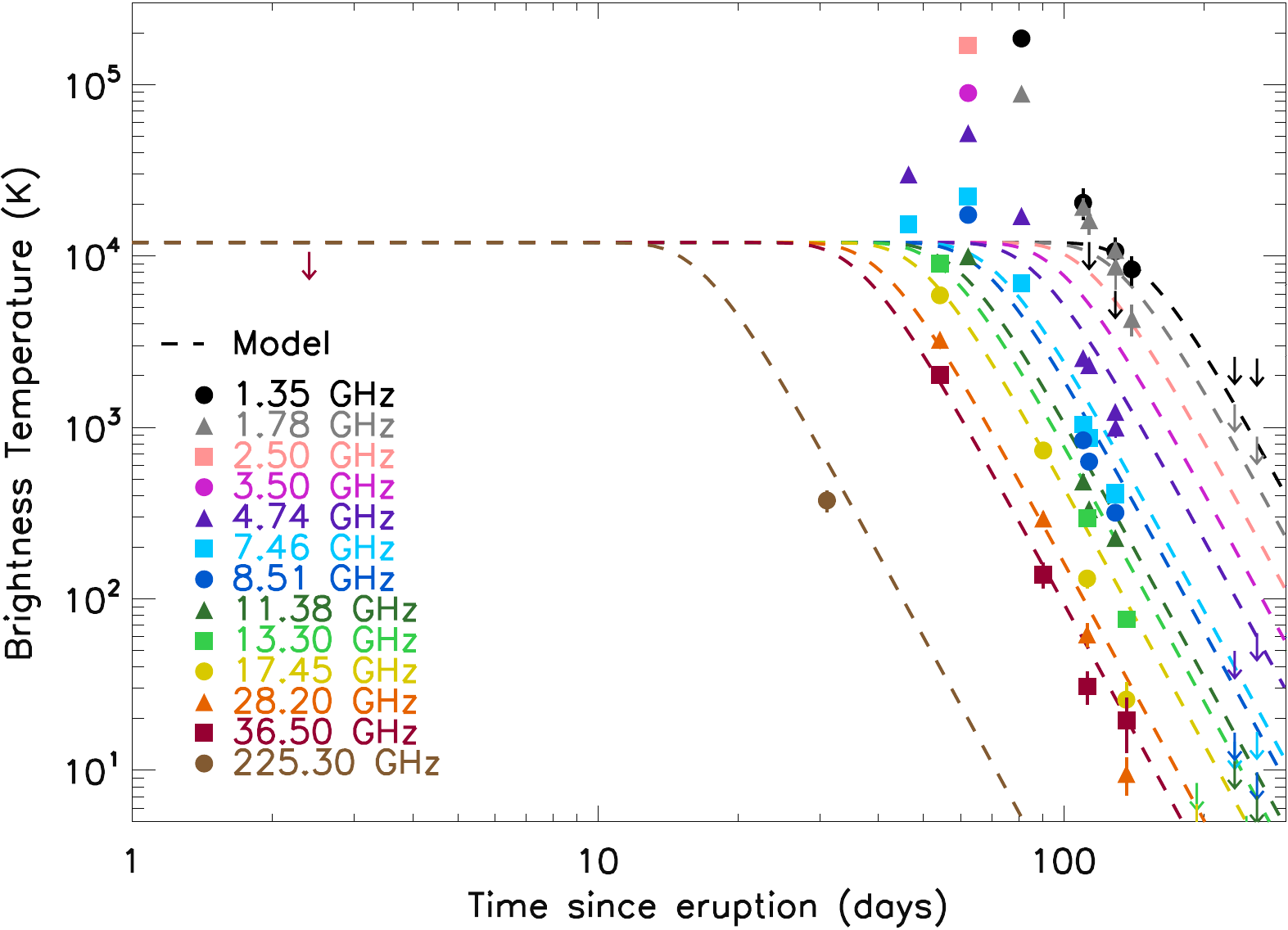}
\label{fig:Flux} 
\end{minipage}
\end{figure*}

Moving forward with the test of the Hubble-flow model, we fit the data
at all 
frequencies  
simultaneously to minimize the 
$\chi^2$ value, weighted by total measurement and systematic
errors. To obtain our fit, we fixed the distance to 4~kpc, the maximum
ejecta velocity to $\v_{HWZI} = 4000$~km~s$^{-1}$, and the ejection
time to $t_0=56038.0$~MJD, allowing the temperature $T$, the mass
ejected M$_{ej}$, and the ratio between {minimum and maximum} velocity,
$\zeta$, to vary. The best-fit parameters for these values were ${\rm
  T} = 1.2\times 10^{4}$~K, M$_{ej} = 2.6 \times 10^{-5}$ M$_{\odot}$,
and $\zeta {\equiv} \frac{\v_{min}}{\v_{HWZI}} = 0.84$.
The fit appears roughly consistent with the data during the decline
portion of the radio light curve, after 
approximately day 100 (see Figure~\ref{fig:Flux}, top panel).
At the lowest frequencies, however, {the data deviate strongly} from the
model (see Figure~\ref{fig:Flux}, middle panel), with observed flux density an order of magnitude greater than the model around the radio peak. The $\chi^2_r$ of 387 indicates that the overall fit is unacceptable. 
We refer to the emission that dominated at low frequencies before day
100, and that deviated strongly from the Hubble-flow expectations, as the
{\em radio flare}. Both the radio flare and the radio emission from the photoionized flow
peaked around day 60.

A comparison among the brightness temperatures for the Hubble-flow model
and for our observations reveals the character of the deviation from
the Hubble-flow model. 
{At radio frequencies, the brightness temperature of an expanding shell is:}
\be
\label{eq:temp}
T_b(\nu, t) \sim \frac{S_\nu (t)c^2 D^2}{2 \pi k_b \nu^2 ({\v_{ej} }\,t)^2},
\ee

\noindent where D is the distance to the source, $c$ is the speed of
light, $k_b$ is Boltzmann's constant, and ${\v_{ej}}${is the velocity of the ejecta} \citep[e.g.,][]{Bode08}.  {This approximation assumes a spherical geometry for the nova shell; 
however, it remains a useful measure even for non-spherical geometries (see discussion in section \ref{subsec:gamma}).}
Brightness temperature {acts as} a measure of surface
brightness, {giving} a lower limit to the physical temperature of
thermally emitting material.  For freely expanding spherical ejecta
with ${\rm T} = 1.2\times 10^{4}$~K, $T_b$ would remain at the physical
temperature of the ejecta while they were optical thick, and then drop
as the ejecta became optically thin (as shown by the dashed lines in
Figure~\ref{fig:Flux} bottom panel).  In contrast to expectations for freely
expanding ejecta with a constant temperature, the brightness
temperature rose dramatically to a peak above $T_b \approx 10^5$~K at
low frequencies 60 to 80 days after the start of the eruption (see Figure~\ref{fig:Flux} bottom panel). 
At high frequencies and after about day 100, the Hubble-flow
model appears 
consistent the data. 

\subsection{X-ray emission}
\indent

The X-ray spectra obtained through day 
53 are reasonably well described by
an absorbed thermal plasma model ({\tt tbabs*apec} in Xspec), with a
temperature that decreased over time.
To determine parameters associated with the X-ray emission, we modelled
each spectrum in Xspec v.12.8.2, obtaining only limits on parameters
for many of the 
observations due to the small number of detected counts.
We fixed the elemental abundances of the plasma to the solar values of
\citet{Wilms00}, which is likely an over-simplification,
as most novae are observed to have highly non-solar abundances.  The
absorbing column constraints in all observations are consistent with
absorption due to the interstellar medium only -- we find no evidence
of the highly absorbed X-ray emission found in other novae at early
times \citep[e.g., V382 Vel;][]{Mukai01}. Based on the fits to the
longer observations on days 19 and 49/50, we see clear
evidence for a large drop in the plasma temperature, from $>$33~keV to
$1.3 \pm 0.1$~keV.

Starting with the day 64.5 spectrum, we find a rise in counts at
energies lower than 1.0 keV, consistent with the appearance of {a}
supersoft source at that time.  Fits to these spectra with a single
absorbed-thermal plasma are poor, 
and also result in much larger values of N(H) than observed in earlier
spectra.  It is likely that this rise in soft flux is due to the
emergence of the supersoft emission from the still-burning white dwarf
photosphere.  Adding a blackbody component to the model does improve
the fit, but the low number of counts mean that the parameters are
unconstrained in most of the observations.  In the observation on day
68, sufficient counts were obtained to result in a constrained model
fit, with best fitting blackbody temperature of 50 (+36,-18) eV.  The
normalization of the blackbody (0.0065 +0.06, -0.004) implies a
luminosity in the range 3 x 10$^{35}$ to 10$^{37}$~erg~s$^{-1}$,
typical of emission-line dominated supersoft sources \citep{Ness13}.
Given our estimate of the ejecta mass from fits to the thermal
component of the radio emission, and the velocity of the ejecta (from
optical line profiles), it is
reasonable that the ejecta became diffuse enough to reveal the
supersoft X-ray emission from the surface of the WD at around day 60. 

%
%


\section{Discussion and Interpretation}

\subsection{Evidence for non-thermal emission}
\indent

\indent

The strongest radio emission from V5589~Sgr at low frequencies -- which we term the {\em
  radio flare} -- does not appear to have been produced by unshocked, isothermal
ejecta.  The properties of the radio-flare
emission deviate strongly 
from the 
Hubble-flow model, which was proposed to reproduce the behaviour of
unshocked ejecta \citep{Seaquist77, Hjellming79}.  In addition, the
maximum $T_b$ 
for V5589~Sgr, which occurred during the radio flare, was higher than
expected for a photoionized outflow 
\citep{Cunningham15}.  For the high radio flux densities at low
frequencies on days 62 and 81 (MJD 56100.3 and 56119.1) to have been
thermal, the emitting material would have required an electron
temperature 
well in excess of the brightness temperature $T_b \sim 10^5$~K (as
measured at 2.5~GHz on day 62 and at 1.35~GHz on day 81).  Ejecta with
$T \approx 10^5$~K could in principle have produced the emission
if the ejecta were optically thick (such that the radio
photosphere closely traced the outermost edge of the ejecta) at low
frequencies.  However, the very flat radio spectra -- $\alpha = 0.14
\pm 0.04$ on day 62.3 and $\alpha = 0.16 \pm 0.04$ on day 81.1 -- are
inconsistent with optically thick thermal emission.  Therefore, if the
radio-flare emission was thermal, the emitting material most likely
had a physical 
temperature significantly greater than $10^5$~K.  Unshocked ejecta,
however, are typically observed to have $T \sim 10^4$~K
\citep{Seaquist77, Hjellming79, Nelson14, Weston16}.  Moreover,
\citet{Cunningham15} argued on theoretical grounds that
photoionziation heating of ejecta by the residual burning of nuclear
fuel in a shell on the surface of the WD leads to temperatures of at most
a few times $10^4$~K, even for photoionization by hot, luminous
high-mass WDs \citep[T. Cunningham, private communication; see also Table~2
of][]{Cunningham15}. Given that the temperature of any thermally
emitting material would have had to have been at least order of
magnitude greater 
than expected for photoionized ejecta (for a distance
of 4~kpc), we conclude that photoionized flows were insufficient to
generate the radio flare.


The strength, spectrum, and rapid rise of the radio flare from V5589~Sgr instead imply that it was produced by shocks.  Shocks can both heat ejecta to temperatures of around $10^6$~K and accelerate particles that subsequently produce synchrotron emission, as in the $\gamma$-ray bright classical nova V959~Mon \citep{Chomiuk14}, the early radio flare from classical-nova V1723~Aql \citep{Weston16}, and as is well established in supernovae \citep[e.g.,][]{Chevalier82,Weiler02, Chandra15}.
In V5589~Sgr, the X-ray emission from plasma with $T > 10^6$~K that
$Swift$/XRT detected between days 19 and 68 (see
Table~\ref{tab:Xrays}) confirms that colliding flows generated shocks
in V5589~Sgr.


Comparing the radio and X-ray fluxes approximately two months after
the start of the outburst, however, shows that shocks probably did not
heat enough material to explain the radio flare as bremsstrahlung from
hot gas. In particular, the emission measure of the X-ray emitting gas
($EM_x = \int n_e^2 dV$, where $n_e$ is the electron density and $V$
is the emitting volume) was several orders of magnitude too small for
the hot ($T > 10^6$~K) plasma to have produced the observed radio
emission between day 60 and day 80, independent of distance.  Although
$EM_x$ on day 65 was quite uncertain due to the short duration of the
$Swift$/XRT observation on that day, it had remained approximately
constant between day 19 and day 53 at a few times $10^{56}\;\Big(\frac{d}{4\,{\rm
  kpc}}\Big)^2$~cm$^{-3}$.  Given the continuous expansion of the remnant,
and presumed subsequent gradual decrease in density of shocked
material, $EM_x$ is unlikely to have suddenly increased between day 53
and 65.  In contrast, an X-ray emission measure on the order of
$10^{60}\;\Big(\frac{d}{4\,{\rm
  kpc}}\Big)^2$~cm$^{-3}$ would have been needed to
explain the radio brightness (assuming that the hot gas was situated
in a thin enough shell to justify having an approximately constant
density profile).  The amount of $T > 10^6$~K gas was therefore too
small to explain the radio flare by more than three orders of
magnitude. This finding is {\em independent} of the distance to
  V5589~Sgr. Moreover, although the peak radio brightness temperature
could in principle have been generated by $10^{-5}~\Msun$ of warm
($10^5$~K) gas,  
the SMARTS optical spectra do not provide any evidence for such a
large quantity of warm gas.   
In addition, that $EM_x$ stayed roughly constant (even while the
density was almost certainly dropping) shows that large quantities of
shock-heated gas were not being cooled out of the X-ray regime to $T
\sim 10^5$~K. It is therefore difficult to reconcile the radio flare
with bremsstrahlung from shock-heated hot plasma.

With not enough warm ($T \sim 10^5$~K) or hot ($T \sim 10^6$~K) gas to
produce the observed peak radio flux, we propose that non-thermal,
synchrotron emission from particles accelerated in shocks dominated
the radio flare.  One concern about this proposal is that the radio
emission near the peak of the radio flare did not have the spectral
index of $\alpha =-0.7$ typically associated with synchrotron emission
\citep{Chevalier82, Weiler02}.  \citet{Weston16} encountered this same issue when
interpreting the early-time flare from the V1723~Aql.  They argued
that the radio spectrum near the peak of the flare in that source
could have been affected by free-free absorption.  Given that we
expect plenty of $T \sim 10^4$~K gas to be present in the ejecta, and
that we expect shocks themselves to photoionize upstream material
\citep{Metzger14},  it is plausible that free-free absorption could
also be affecting the radio spectrum near the peak of the flare in
V5589~Sgr.  Because the hot, X-ray emitting gas was insufficient to
produce the radio flare and free-free absorption is plausible, we
consider synchrotron emission to be the most likely explanation for
the radio flare despite the slightly rising radio spectrum.


The rapid radio brightening, or flare, 
between days 46.2 and 62.3 also provides evidence for shocks as the
origin of the strongest low-frequency radio emission from V5589~Sgr.  At both 5.0 and 6.8~GHz, the radio flux densities increased much faster than possible for a freely expanding, $T\sim 10^4$~K (i.e., photoionized) remnant ejected at the start of the eruption.  Whereas a freely expanding, constant-temperature flow can produce radio flux densities that increase as fast as $t^2$ (for an optically thick spherical outflow), the 5.0-GHz flux density rose as $t^{3.9}$, even while the radio spectra showed that the emitting material were not optically thick.  For the radio flare to have been due to freely expanding ejecta with $T\sim 10^4$~K, the emitting material would have had to have been ejected not at the start of the eruption, but more than 26 days {\em after} $t_0$.  Neither the optical spectra \citep{Walter12} nor the optical light curve \citep{Mroz15}, however, show evidence for a strong new outflow a month into the eruption.  Furthermore, the contrast between the fairly constant X-ray emission measure, $EM_x$, and the rapidly rising radio flux between 46.2 and 62.3 days after the start of the eruption further supports our contention that the radio flare was not thermal emission from shock-heated gas. Like our analysis of the emission measure, this result is also independent of distance. The appearance of a new source of radio emission on the order of months after the start of the eruption was more consistent with the synchrotron knots tracing shocks in V959~Mon, at least one of which was still brightening months after the detection of $\gamma$-rays \citep{Chomiuk14}.
%
%
In the speed of its radio brightening, the radio flare from V5589~Sgr was very similar to the shock-powered early flare in V1723~Aql \citep{Weston16}.

\subsection{Connection to $\gamma$-ray producing novae}
\label{subsec:gamma}
\indent


The similar viewing angles of V5589~Sgr and the $\gamma$-ray bright
novae V959~Mon enable us to speculate about the velocity and structure
of the colliding flows from V5589~Sgr. Because light curves of both
V959~Mon and V5589~Sgr show eclipses \citep{Page13, Mroz15}, both
binaries have inclinations that are close to $90^\circ$, so that the
orbital planes are observed nearly edge on. In V959~Mon, this viewing
angle led to a horn-shaped component in optical emission line profiles
\citep[e.g., see Fig.~4 of][]{Ribeiro13} -- from what radio and HST
observations revealed to be a slow, dense inner flow (\citealt{Chomiuk14}; {Sokoloski et al. in preparation}).  
This slow flow in V959~Mon consisted of a spherical shell-like
structure  with a strong equatorial density enhancement
(\citealt{Chomiuk14}; {Sokoloski et al. in preparation}). Because the optical emission lines in
SMARTS spectra of V5589~Sgr contain a similar horn-shaped component,
with peaks near Doppler shifts of approximately $\pm
1,800$~km~s$^{-1}$ 
\citep[e.g., in H$\alpha$ on MJD 56048.4, 56101.0, and 56103.0, and in OIII on MJD 56057.1;][see Figure~\ref{fig:OpticalSpectra}]{Walter12}, 
 we hypothesize that the remnant around V5589~Sgr also included an
 inner shell-like structure expanding at a speed of around
 $1,800$~km~s$^{-1}$, with a strong equatorial density
 enhancement\footnote{If such an inner, slow structure with an
   equatorial density enhancement is common in novae, that could
   explain the narrow cores in the optical emission lines of some
   novae that are viewed {more} pole on, such as {V339 Del} \citep{Schaefer14}.
   In those systems, the face-on equatorial torus would generate an
   emission component with very low radial velocities.}.  In V959~Mon,
 the slow equatorial torus collimated a faster flow into a bi-conical
 shape (\citealt{Ribeiro13, Shore13, Chomiuk14}; {Sokoloski et al. in preparation}), which
 produced `shoulders' in the optical line profiles that were similar
 to the shoulders out to radial velocities of $\pm 4,000$~km~s$^{-1}$
 in V5589~Sgr.  

In V959~Mon, projection effects caused the shoulders in the optical emission lines to have velocity widths that were less than the the maximum speed of the bi-conical fast flow.  For V5589~Sgr, the high temperature of the X-ray emitting plasma on day 19 ($T > 3 \times 10^8$~K) reveals that the difference between the speeds of the fast and slow flows was at least initially greater than 3,800~km~s$^{-1}$ (using the expression
$T_{ps}=\frac{3}{16}\frac{\mu m_p}{k_B} \v_s^2$
for the post-shock temperature, and assuming that the shock speed was
$\frac{4}{3}$ times the difference between the speeds of the fast and
slow flows). The fast flow thus probably had a maximum velocity of
more than 5,000~km~s$^{-1}$.  In fact, faint extended H$\alpha$ wings
in the early spectra from SMARTS, as well as spectroscopy in the first
few days of eruption from the Crimean Laboratory, Moscow State
University \citep{ATEL4094} support the existence of flows with speeds
in excess of 4,000~km~s$^{-1}$.  
That the core of the optical emission lines only extended out to
4,000~km~s$^{-1}$ is thus consistent with a bi-conical outflow, as in
V959~Mon. The optical, radio, and X-ray observations are therefore
consistent with V5589~Sgr having a similar ring$+$bipolar structure as
V959~Mon.




Our previous arguments regarding the origin of the radio flare
remain valid under this slightly refined picture of the structure of
the V5589~Sgr remnant.  For example, in our calculation of the
brightness temperature in equation~\ref{eq:temp}, { }the maximum
outflow velocity of the ejecta {was} probably actually greater than
$4,000$~km~s$^{-1}$. {However,} the area of the bi-conical
remnant on the sky is smaller than in the case of the spherical
remnant (as long as the half-opening angles for the cones are wider
than 45$^\circ$), making {the value of} $T_b$ even higher than in the spherical
case. Thus, the radio flare is still more consistent with being
shock powered than being bremsstrahlung from unshocked, photoionized
ejecta. 

%
%


A torus of relatively slow-moving material 
in the equatorial plane
may originate in one of several
ways. As suggested by \citet{Chomiuk14} for V959 Mon, the
development of a dense toroidal outflow is consistent with
hydrodynamic simulations of how the binary companion might 
influence the behaviour of nova ejecta
\citep{Livio90,Lloyd97,Chomiuk14}. Therefore, interactions between the
companion and the common envelope may lead to a more dense
concentration of material in the orbital plane of the
binary. Alternatively, there may be a pre-existing circumbinary
material stemming from ejections of material from
the accretion disk
\citep[e.g.,][]{Sytov07}. 
Rings also may form in the nebulae of post-main sequence rotating
stars from Eddington-luminosity driven outflow, due to the
interaction of stellar winds \citep[e.g.,][]{Chita08}. 
Although the mass in
the {accretion} disk could 
be significant for very wide binaries with red-giant donors, it is not
likely to be high enough in V5589~Sgr to generate an 
equatorial torus with mass comparable to that of the rest of the
ejecta. 
No matter what their origin, equatorial density enhancements in the
remnants of novae provide important constraints on either the mass
transfer before the eruption or the ejection mechanism for the nova. 


The radio emission from V5589~Sgr was probably dominated by a shock-powered flare rather than the more usual case of bremsstrahlung from the $T \sim 10^4$~K photoionized remnant because of the high speed and relatively low mass of the ejecta.  The 2010 eruption of V1723~Aql first produced a shock-powered radio flare and then a period of strong thermal radio emission from a $T \sim 10^4$~K expanding remnant with a mass of $2 \times 10^{-4}$ M$_{\odot}$ and maximum velocity of 1500 km s$^{-1}$; both the flare and the later peak in the radio light curve reached approximately the same brightness level \citep{Weston16}.  But the expected peak radio brightness from photoionized ejecta is a strong function of the ejecta mass and the speed of the outflow \citep{Bode08}, both in the sense that would diminish the expected radio emission from photoionized ejecta in V5589~Sgr.  If the strength of shock-powered synchrotron emission is less sensitive to these parameters, or enhanced in a high-speed shock, then the shock-powered flare could become the dominant element of the radio light curve.


Interestingly, despite the importance of shocks and possibly even
relativistic particles in generating its radio emission, V5589~Sgr was
less bright in $\gamma$-rays than several other classical novae at
similar distances. 
Since the $Fermi$/LAT sensitivity for the location of V5589~Sgr
($l=5.0$, $b=+3.1$) is comparable to that for V1324~Sco (l=357.4,
b=-2.9), the daily GeV flux of V5589~Sgr almost certainly stayed below
$5 \times 10^{-7}$~photons~cm$^{-2}$~s$^{-1}$, which is below the
level seen in  
V1324~Sco and V959 Mon \citep{Ackermann14}.  
It remains to be seen how the upper limit on the $\gamma$-ray flux
from V5589~Sgr compares with the detections of the four later
Fermi-detected novae, all of which were observed in pointed mode with
fluxes as faint as 2--3$\times
10^{-7}$~photons~cm$^{-2}$~s$^{-1}$. When the distances of
$1.4\pm0.04$~kpc for V959~Mon \citep{Linford15}, $>6.5$~kpc for
V1324~Sco \citep{Finzell15}, and $4.5\pm0.6$~kpc for V339~Del
\citep{Schaefer14} are taken into account, V5589~Sgr appears to have
been less luminous in the $\gamma$-rays than at least V1324~Sco and
V339~Del. Since V339~Del is thought to have a low inclination and
therefore to be viewed pole on \citep{Schaefer14}, viewing angle could
play a role in setting the $\gamma$-ray luminosity. Alternatively,
whereas \citet{Metzger15} argued that most relativistic protons in
V1324~Sco and V339~Del were trapped and 
collided with proton
targets to produce $\gamma$-rays, the low X-ray absorbing column for V5589~Sgr
suggests that the ejecta mass might have been too low and/or the
expansion speeds 
too high for this to have been the case for V5589~Sgr.

 {The presence of high energy shocks and} ${\gamma}${-rays from novae is still an active area of research with many complexities and open questions. Shocks themselves may be apparently absent in systems where expected to appear, such as in the symbiotic Nova Sco 2014} \citep{Joshi15}.  {The speed of the shocks and the angle at which the shocks are viewed may play a large part in determining whether} ${\gamma}${-rays are detected in nearby novae} \citep{Metzger15, DmitrievichVlasov16}. {If this were the case, it could explain systems such as the clearly shock-powered yet } ${\gamma}${-ray faint V5589 Sgr.}




Strengthening the connection between V5589~Sgr and classical novae
that have been detected in the $\gamma$-rays, 
V5589~Sgr is not
the only nova to show a notable excess of radio flux at lower
frequencies. The 1.75 GHz flux
density from $Fermi$-detected nova V959~Mon peaked around 170 days
after the start of 
that eruption, counter to expectations based on models of 
expanding thermal ejecta. This excess is pictured in Chomiuk et
al.~2014 (Extended Data Figure 4); while none of the light curves at
frequencies 
spanning 
7.4--225 GHz are very well fit by a simple Hubble flow model, by far
the most divergent is the lowest frequency light curve, which shows a
peak flux density of 11.2 mJy at 1.75 GHz, when a peak of just $\sim$4.5
mJy is predicted by the Hubble flow model.  As in the case of
V5589~Sgr, this lower-frequency radio peak in V959~Mon
occurred 
around the same time as the maxima at higher frequencies ($\sim$30
GHz). High-resolution VLBI images of V959~Mon confirm the presence of
synchrotron-emitting shocks in that nova.  So it is likely that the
radio spectral evolution of both V959~Mon and V5589~Sgr can be
explained by the same type of model.

\section{Conclusions}
\indent

Despite not being detected in the $\gamma$-rays, V5589~Sgr adds to the growing body of evidence that shocks play an important role in normal novae, and that these shocks can accelerate particles to relativistic speeds.  In particular,

\begin{itemize}

\item The low-frequency radio emission from V5589~Sgr during its 2012
  eruption was dominated by a flare that appears to have been too
  strong to have been due to bremsstrahlung from photoionized ejecta.
  The speed with which the radio flare appeared and its radio spectral
  evolution, 
{are} inconsistent with expectations {for} freely
  expanding, photoionized ejecta. We conclude that the radio flare
  from V5589~Sgr was powered by shocks.

\item Hard ($kT > 1$~keV) X-ray emission confirms that shocks were
  present, but that they did not heat enough plasma for the radio flare two
  months after the start of the eruption to have been due to thermal
  bremsstrahlung from hot, $T > 10^6$~K shock-heated gas. 

\item Synchrotron emission from relativistic particles accelerated in
  shocks is {the most natural alternative mechanism for the} radio flare.   

\item V5589~Sgr is one of the best examples of shock-powered
  radio emission from a nova without a red-giant donor. Another clear
  example is V1723~Aql.  We expect that other novae with ejecta that
  have high speeds and fairly low masses could also produce strong,
  shock-powered radio emission.

\item Using the similar edge-on inclinations of V5589~Sgr and V959~Mon
  to aid in the interpretation of optical emission-line profiles, and
  the temperature of the shock-heated X-ray emitting plasma to infer
  flow speeds in the shock, we hypothesize that in V5589~Sgr a fast
  flow with a maximum speed of more than 5000~km~s$^{-1}$ collided
  with, and was loosely collimated by, a slower flow 
that was concentrated in the equatorial plane. 

\item The lack of detectable $\gamma$-ray emission from V5589~Sgr,
  despite direct evidence for strong shocks and indirect evidence for
  relativistic particles, hints at the possibility that  
low ejecta densities due to high outflow speeds could have played a
role in V5589~Sgr's low $\gamma$-ray luminosity.    


\end{itemize}
\section{Acknowledgements}
\indent

We thank NRAO for its generous allocation of time which made this work possible. The National Radio Astronomy Observatory is a facility of the National Science Foundation operated under cooperative agreement by Associated Universities, Inc. J. Weston and J.L. Sokoloski acknowledge support from NSF award AST-1211778. J. Weston was supported in part by a Student Observing Support award from NRAO (NRAO 343777). L. Chomiuk, J. Linford, and T. Finzell are supported by NASA Fermi Guest Investigator grant NNH13ZDA001N-FERMI. T. Nelson was supported in part by NASA award NNX13A091G.  F. M. Walter thanks the Provost of Stony Brook University for providing support for continued participation in SMARTS. We acknowledge with thanks the variable star observations from the AAVSO International Database contributed by observers worldwide and used as reference in this work. Thanks to C.C. Cheung for use of data from VLA program S4322. Thanks to Glen Petitpas for SMA data reduction. Thanks to Tim Cunningham, Eric Gotthelf, David Schiminovich, and Slavko Bogdanov for useful discussion.

\bibliography{bib_sgr}

\begin{thebibliography}{}

\bibitem[\protect\citeauthoryear{{Abdo}, {Ackermann}, {Ajello}, {Atwood},
  {Baldini}, {Ballet}, {Barbiellini}, {Bastieri}, {Bechtol}, {Bellazzini} \& et
  al.}{{Abdo} et~al.}{2010}]{Abdo10}
{Abdo} A.~A.,  {Ackermann} M.,  {Ajello} M.,  {Atwood} W.~B.,  {Baldini} L.,
  {Ballet} J.,  {Barbiellini} G.,  {Bastieri} D.,  {Bechtol} K.,  {Bellazzini}
  R.,    et al. 2010, Science, 329, 817

\bibitem[\protect\citeauthoryear{{Acero}, {Ackermann}, {Ajello}, {Albert},
  {Wood}, {Wood} \& {Zimmer}}{{Acero} et~al.}{2015}]{Acero15}
{Acero} F.,  {Ackermann} M.,  {Ajello} M.,  {Albert} A.,  {Wood} K.~S.,  {Wood}
  M.,    {Zimmer} S.,  2015, \apjs, 218, 23

\bibitem[\protect\citeauthoryear{{Ackermann}, {Ajello}, {Albert}, {Baldini},
  {Teyssier} \& {Fermi-LAT Collaboration}}{{Ackermann}
  et~al.}{2014}]{Ackermann14}
{Ackermann} M.,  {Ajello} M.,  {Albert} A.,  {Baldini} L.,  {Teyssier} F.,
  {Fermi-LAT Collaboration} 2014, Science, 345, 554

\bibitem[\protect\citeauthoryear{{Anupama}, {Kamath}, {Ramaprakash},
  {Kantharia}, {Hegde}, {Mohan}, {Kulkarni}, {Bode}, {Eyres}, {Evans} \&
  {O'Brien}}{{Anupama} et~al.}{2013}]{Anupama13}
{Anupama} G.~C.,  {Kamath} U.~S.,  {Ramaprakash} A.~N.,  {Kantharia} N.~G.,
  {Hegde} M.,  {Mohan} V.,  {Kulkarni} M.,  {Bode} M.~F.,  {Eyres} S.~P.~S.,
  {Evans} A.,    {O'Brien} T.~J.,  2013, \aap, 559, A121

\bibitem[\protect\citeauthoryear{{Banerjee}, {Joshi}, {Venkataraman}, {Ashok},
  {Marion}, {Hsiao} \& {Raj}}{{Banerjee} et~al.}{2014}]{Banerjee14}
{Banerjee} D.~P.~K.,  {Joshi} V.,  {Venkataraman} V.,  {Ashok} N.~M.,  {Marion}
  G.~H.,  {Hsiao} E.~Y.,    {Raj} A.,  2014, \apjl, 785, L11

\bibitem[\protect\citeauthoryear{{Bode} \& {Evans}}{{Bode} \&
  {Evans}}{2008}]{Bode08}
{Bode} M.~F.,  {Evans} A.,  2008, {Classical Novae}.
Cambridge University Press

\bibitem[\protect\citeauthoryear{{Bode}, {O'Brien}, {Osborne}, {Page},
  {Senziani}, {Skinner}, {Starrfield}, {Ness}, {Drake}, {Schwarz}, {Beardmore},
  {Darnley}, {Eyres}, {Evans}, {Gehrels}, {Goad}, {Jean}, {Krautter} \&
  {Novara}}{{Bode} et~al.}{2006}]{Bode06}
{Bode} M.~F.,  {O'Brien} T.~J.,  {Osborne} J.~P.,  {Page} K.~L.,  {Senziani}
  F.,  {Skinner} G.~K.,  {Starrfield} S.,  {Ness} J.-U.,  {Drake} J.~J.,
  {Schwarz} G.,  {Beardmore} A.~P.,  {Darnley} M.~J.,  {Eyres} S.~P.~S.,
  {Evans} A.,  {Gehrels} N.,  {Goad} M.~R.,  {Jean} P.,  {Krautter} J.,
  {Novara} G.,  2006, \apj, 652, 629

\bibitem[\protect\citeauthoryear{{Buil}}{{Buil}}{2012}]{CBET3089Buil}
{Buil} C.,  2012, Central Bureau Electronic Telegrams, 3089, 3

\bibitem[\protect\citeauthoryear{{Cao}, {Kasliwal}, {Neill}, {Kulkarni}, {Lou},
  {Ben-Ami}, {Bloom}, {Cenko}, {Law}, {Nugent}, {Ofek}, {Poznanski} \&
  {Quimby}}{{Cao} et~al.}{2012}]{Cao12}
{Cao} Y.,  {Kasliwal} M.~M.,  {Neill} J.~D.,  {Kulkarni} S.~R.,  {Lou} Y.-Q.,
  {Ben-Ami} S.,  {Bloom} J.~S.,  {Cenko} S.~B.,  {Law} N.~M.,  {Nugent} P.~E.,
  {Ofek} E.~O.,  {Poznanski} D.,    {Quimby} R.~M.,  2012, \apj, 752, 133

\bibitem[\protect\citeauthoryear{{Chandra}, {Chevalier}, {Chugai}, {Fransson}
  \& {Soderberg}}{{Chandra} et~al.}{2015}]{Chandra15}
{Chandra} P.,  {Chevalier} R.~A.,  {Chugai} N.,  {Fransson} C.,    {Soderberg}
  A.~M.,  2015, \apj, 810, 32

\bibitem[\protect\citeauthoryear{{Cheung}, {Donato}, {Wallace}, {Corbet},
  {Dubus}, {Sokolovsky} \& {Takahashi}}{{Cheung} et~al.}{2010}]{Cheung10}
{Cheung} C.~C.,  {Donato} D.,  {Wallace} E.,  {Corbet} R.,  {Dubus} G.,
  {Sokolovsky} K.,    {Takahashi} H.,  2010, The Astronomer's Telegram, 2487, 1

\bibitem[\protect\citeauthoryear{{Cheung} C.~C.}{{Cheung}}{2013}]{Cheung13}
{Cheung} C.~C. o.,  2013, arXiv:1304.3475

\bibitem[\protect\citeauthoryear{{Chevalier}}{{Chevalier}}{1982}]{Chevalier82}
{Chevalier} R.~A.,  1982, \apj, 259, 302

\bibitem[\protect\citeauthoryear{{Chita}, {Langer}, {van Marle},
  {Garc{\'{\i}}a-Segura} \& {Heger}}{{Chita} et~al.}{2008}]{Chita08}
{Chita} S.~M.,  {Langer} N.,  {van Marle} A.~J.,  {Garc{\'{\i}}a-Segura} G.,
  {Heger} A.,  2008, \aap, 488, L37

\bibitem[\protect\citeauthoryear{{Chomiuk}, {Krauss}, {Rupen}, {Nelson}, {Roy},
  {Sokoloski}, {Mukai}, {Munari}, {Mioduszewski}, {Weston}, {O'Brien}, {Eyres}
  \& {Bode}}{{Chomiuk} et~al.}{2012}]{Chomiuk12}
{Chomiuk} L.,  {Krauss} M.~I.,  {Rupen} M.~P.,  {Nelson} T.,  {Roy} N.,
  {Sokoloski} J.~L.,  {Mukai} K.,  {Munari} U.,  {Mioduszewski} A.,  {Weston}
  J.,  {O'Brien} T.~J.,  {Eyres} S.~P.~S.,    {Bode} M.~F.,  2012, \apj, 761,
  173

\bibitem[\protect\citeauthoryear{{Chomiuk}, {Linford}, {Finzell}, {Sokoloski},
  {Weston}, {Zheng}, {Nelson}, {Mukai}, {Rupen} \& {Mioduszewski}}{{Chomiuk}
  et~al.}{2013}]{ATEL5382}
{Chomiuk} L.,  {Linford} J.,  {Finzell} T.,  {Sokoloski} J.,  {Weston} J.,
  {Zheng} Y.,  {Nelson} T.,  {Mukai} K.,  {Rupen} M.,    {Mioduszewski} A.,
  2013, The Astronomer's Telegram, 5382, 1

\bibitem[\protect\citeauthoryear{{Chomiuk}, {Linford}, {Yang}, {O'Brien},
  {Paragi}, {Mioduszewski}, {Beswick}, {Cheung}, {Mukai}, {Nelson}, {Ribeiro},
  {Rupen}, {Sokoloski}, {Weston}, {Zheng}, {Bode}, {Eyres}, {Roy} \&
  {Taylor}}{{Chomiuk} et~al.}{2014}]{Chomiuk14}
{Chomiuk} L.,  {Linford} J.~D.,  {Yang} J.,  {O'Brien} T.~J.,  {Paragi} Z.,
  {Mioduszewski} A.~J.,  {Beswick} R.~J.,  {Cheung} C.~C.,  {Mukai} K.,
  {Nelson} T.,  {Ribeiro} V.~A.~R.~M.,  {Rupen} M.~P.,  {Sokoloski} J.~L.,
  {Weston} J.,  {Zheng} Y.,  {Bode} M.~F.,  {Eyres} S.,  {Roy} N.,    {Taylor}
  G.~B.,  2014, \nat, 514, 339

\bibitem[\protect\citeauthoryear{{Chomiuk}, {Nelson}, {Mukai}, {Sokoloski},
  {Rupen}, {Page}, {Osborne}, {Kuulkers}, {Mioduszewski}, {Roy}, {Weston} \&
  {Krauss}}{{Chomiuk} et~al.}{2014}]{Chomiuk14b}
{Chomiuk} L.,  {Nelson} T.,  {Mukai} K.,  {Sokoloski} J.~L.,  {Rupen} M.~P.,
  {Page} K.~L.,  {Osborne} J.~P.,  {Kuulkers} E.,  {Mioduszewski} A.~J.,  {Roy}
  N.,  {Weston} J.,    {Krauss} M.~I.,  2014, \apj, 788, 130

\bibitem[\protect\citeauthoryear{{Cunningham}, {Wolf} \&
  {Bildsten}}{{Cunningham} et~al.}{2015}]{Cunningham15}
{Cunningham} T.,  {Wolf} W.~M.,    {Bildsten} L.,  2015, \apj, 803, 76

\bibitem[\protect\citeauthoryear{{Darnley}, {Ribeiro}, {Bode}, {Hounsell} \&
  {Williams}}{{Darnley} et~al.}{2012}]{Darnley12}
{Darnley} M.~J.,  {Ribeiro} V.~A.~R.~M.,  {Bode} M.~F.,  {Hounsell} R.~A.,
  {Williams} R.~P.,  2012, \apj, 746, 61

\bibitem[\protect\citeauthoryear{{Das}, {Banerjee} \& {Ashok}}{{Das}
  et~al.}{2006}]{Das06}
{Das} R.,  {Banerjee} D.~P.~K.,    {Ashok} N.~M.,  2006, \apjl, 653, L141

\bibitem[\protect\citeauthoryear{{Dmitrievich Vlasov}, {Vurm} \&
  {Metzger}}{{Dmitrievich Vlasov} et~al.}{2016}]{DmitrievichVlasov16}
{Dmitrievich Vlasov} A.,  {Vurm} I.,    {Metzger} B.~D.,  2016, ArXiv:
  1603.05194

\bibitem[\protect\citeauthoryear{{Esipov}, {Sokolovsky} \& {Korotkiy}}{{Esipov}
  et~al.}{2012}]{ATEL4094}
{Esipov} V.~F.,  {Sokolovsky} K.~V.,    {Korotkiy} S.~A.,  2012, The
  Astronomer's Telegram, 4094, 1

\bibitem[\protect\citeauthoryear{{Eyres}, {Bode}, {O'Brien}, {Watson} \&
  {Davis}}{{Eyres} et~al.}{2000}]{Eyres00}
{Eyres} S.~P.~S.,  {Bode} M.~F.,  {O'Brien} T.~J.,  {Watson} S.~K.,    {Davis}
  R.~J.,  2000, \mnras, 318, 1086

\bibitem[\protect\citeauthoryear{{Finzell}, {Chomiuk}, {Munari} \&
  {Walter}}{{Finzell} et~al.}{2015}]{Finzell15}
{Finzell} T.,  {Chomiuk} L.,  {Munari} U.,    {Walter} F.~M.,  2015, \apj, 809,
  160

\bibitem[\protect\citeauthoryear{{Friedman}, {York}, {McCall}, {Dahlstrom},
  {Sonnentrucker}, {Welty}, {Drosback}, {Hobbs}, {Rachford} \&
  {Snow}}{{Friedman} et~al.}{2011}]{Friedman11}
{Friedman} S.~D.,  {York} D.~G.,  {McCall} B.~J.,  {Dahlstrom} J.,
  {Sonnentrucker} P.,  {Welty} D.~E.,  {Drosback} M.~M.,  {Hobbs} L.~M.,
  {Rachford} B.~L.,    {Snow} T.~P.,  2011, \apj, 727, 33

\bibitem[\protect\citeauthoryear{{Green}, {Schlafly}, {Finkbeiner}, {Rix},
  {Martin}, {Burgett}, {Draper}, {Flewelling}, {Hodapp}, {Kaiser}, {Kudritzki},
  {Magnier}, {Metcalfe}, {Price}, {Tonry} \& {Wainscoat}}{{Green}
  et~al.}{2015}]{Green15}
{Green} G.~M.,  {Schlafly} E.~F.,  {Finkbeiner} D.~P.,  {Rix} H.-W.,  {Martin}
  N.,  {Burgett} W.,  {Draper} P.~W.,  {Flewelling} H.,  {Hodapp} K.,  {Kaiser}
  N.,  {Kudritzki} R.~P.,  {Magnier} E.,  {Metcalfe} N.,  {Price} P.,  {Tonry}
  J.,    {Wainscoat} R.,  2015, \apj, 810, 25

\bibitem[\protect\citeauthoryear{{Hjellming}, {Wade}, {Vandenberg} \&
  {Newell}}{{Hjellming} et~al.}{1979}]{Hjellming79}
{Hjellming} R.~M.,  {Wade} C.~M.,  {Vandenberg} N.~R.,    {Newell} R.~T.,
  1979, ApJ, 84, 1619

\bibitem[\protect\citeauthoryear{{H{\"o}gbom}}{{H{\"o}gbom}}{1974}]{Hogbom74}
{H{\"o}gbom} J.~A.,  1974, \aaps, 15, 417

\bibitem[\protect\citeauthoryear{{Joshi}, {Banerjee}, {Ashok}, {Venkataraman}
  \& {Walter}}{{Joshi} et~al.}{2015}]{Joshi15}
{Joshi} V.,  {Banerjee} D.~P.~K.,  {Ashok} N.~M.,  {Venkataraman} V.,
  {Walter} F.~M.,  2015, \mnras, 452, 3696

\bibitem[\protect\citeauthoryear{{Korotkiy}, {Sokolovsky}, {Brown}, {Gao},
  {Seach}, {Kiytota}, {Guido}, {Howes}, {Sostero}, {Elenin}, {Molotov}, {Koff},
  {Nissinen}, {Nishiyama}, {Kabashima}, {Koberger}, {Vollmann} \&
  {Gerke}}{{Korotkiy} et~al.}{2012}]{CBET3089Korotkiy}
{Korotkiy} S.,  {Sokolovsky} K.,  {Brown} N.~J.,  {Gao} R.~J.,  {Seach} J.,
  {Kiytota} S.,  {Guido} E.,  {Howes} N.,  {Sostero} G.,  {Elenin} L.,
  {Molotov} I.,  {Koff} R.~A.~A.,  {Nissinen} M.,  {Nishiyama} K.,  {Kabashima}
  F.,  {Koberger} H.,  {Vollmann} W.,    {Gerke} V.,  2012, Central Bureau
  Electronic Telegrams, 3089, 1

\bibitem[\protect\citeauthoryear{{Kraft}, {Burrows} \& {Nousek}}{{Kraft}
  et~al.}{1991}]{Kraft91}
{Kraft} R.~P.,  {Burrows} D.~N.,    {Nousek} J.~A.,  1991, \apj, 374, 344

\bibitem[\protect\citeauthoryear{{Krauss}, {Chomiuk}, {Rupen}, {Roy},
  {Mioduszewski}, {Sokoloski}, {Nelson}, {Mukai}, {Bode}, {Eyres} \&
  {O'Brien}}{{Krauss} et~al.}{2011}]{Krauss11}
{Krauss} M.~I.,  {Chomiuk} L.,  {Rupen} M.,  {Roy} N.,  {Mioduszewski} A.~J.,
  {Sokoloski} J.~L.,  {Nelson} T.,  {Mukai} K.,  {Bode} M.~F.,  {Eyres}
  S.~P.~S.,    {O'Brien} T.~J.,  2011, ApJ, 739, L6

\bibitem[\protect\citeauthoryear{{Linford}, {Ribeiro}, {Chomiuk}, {Nelson},
  {Sokoloski}, {Rupen}, {Mukai}, {O'Brien}, {Mioduszewski} \&
  {Weston}}{{Linford} et~al.}{2015}]{Linford15}
{Linford} J.~D.,  {Ribeiro} V.~A.~R.~M.,  {Chomiuk} L.,  {Nelson} T.,
  {Sokoloski} J.~L.,  {Rupen} M.~P.,  {Mukai} K.,  {O'Brien} T.~J.,
  {Mioduszewski} A.~J.,    {Weston} J.,  2015, \apj, 805, 136

\bibitem[\protect\citeauthoryear{{Livio}, {Shankar}, {Burkert} \&
  {Truran}}{{Livio} et~al.}{1990}]{Livio90}
{Livio} M.,  {Shankar} A.,  {Burkert} A.,    {Truran} J.~W.,  1990, \apj, 356,
  250

\bibitem[\protect\citeauthoryear{{Lloyd}, {O'Brien} \& {Bode}}{{Lloyd}
  et~al.}{1997}]{Lloyd97}
{Lloyd} H.~M.,  {O'Brien} T.~J.,    {Bode} M.~F.,  1997, \mnras, 284, 137

\bibitem[\protect\citeauthoryear{{McMullin}, {Waters}, {Schiebel}, {Young} \&
  {Golap}}{{McMullin} et~al.}{2007}]{McMullin07}
{McMullin} J.~P.,  {Waters} B.,  {Schiebel} D.,  {Young} W.,    {Golap} K.,
  2007, in {Shaw} R.~A.,  {Hill} F.,   {Bell} D.~J.,  eds, Astronomical Data
  Analysis Software and Systems XVI Vol.~376 of Astronomical Society of the
  Pacific Conference Series, {CASA Architecture and Applications}.
p.~127

\bibitem[\protect\citeauthoryear{{Metzger}, {Finzell}, {Vurm}, {Hasco{\"e}t},
  {Beloborodov} \& {Chomiuk}}{{Metzger} et~al.}{2015}]{Metzger15}
{Metzger} B.~D.,  {Finzell} T.,  {Vurm} I.,  {Hasco{\"e}t} R.,  {Beloborodov}
  A.~M.,    {Chomiuk} L.,  2015, \mnras, 450, 2739

\bibitem[\protect\citeauthoryear{{Metzger}, {Hasco{\"e}t}, {Vurm},
  {Beloborodov}, {Chomiuk}, {Sokoloski} \& {Nelson}}{{Metzger}
  et~al.}{2014}]{Metzger14}
{Metzger} B.~D.,  {Hasco{\"e}t} R.,  {Vurm} I.,  {Beloborodov} A.~M.,
  {Chomiuk} L.,  {Sokoloski} J.~L.,    {Nelson} T.,  2014, \mnras, 442, 713

\bibitem[\protect\citeauthoryear{{Mink}}{{Mink}}{2011}]{Mink11}
{Mink} D.~J.,  2011, in {Evans} I.~N.,  {Accomazzi} A.,  {Mink} D.~J.,   {Rots}
  A.~H.,  eds, Astronomical Data Analysis Software and Systems XX Vol.~442 of
  Astronomical Society of the Pacific Conference Series, {Data Pipelines for
  the TRES Echelle Spectrograph}.
p.~305

\bibitem[\protect\citeauthoryear{{Mr{\'o}z}, {Udalski}, {Poleski},
  {Soszy{\'n}ski}, {Szyma{\'n}ski}, {Pietrzy{\'n}ski}, {Wyrzykowski},
  {Ulaczyk}, {Koz{\l}owski}, {Pietrukowicz} \& {Skowron}}{{Mr{\'o}z}
  et~al.}{2015}]{Mroz15}
{Mr{\'o}z} P.,  {Udalski} A.,  {Poleski} R.,  {Soszy{\'n}ski} I.,
  {Szyma{\'n}ski} M.~K.,  {Pietrzy{\'n}ski} G.,  {Wyrzykowski} {\L}.,
  {Ulaczyk} K.,  {Koz{\l}owski} S.,  {Pietrukowicz} P.,    {Skowron} J.,  2015,
  \apjs, 219, 26

\bibitem[\protect\citeauthoryear{{Mukai} \& {Ishida}}{{Mukai} \&
  {Ishida}}{2001}]{Mukai01}
{Mukai} K.,  {Ishida} M.,  2001, \apj, 551, 1024

\bibitem[\protect\citeauthoryear{{Munari}, {Joshi}, {Ashok}, {Banerjee},
  {Valisa}, {Milani}, {Siviero}, {Dallaporta} \& {Castellani}}{{Munari}
  et~al.}{2011}]{Munari11}
{Munari} U.,  {Joshi} V.~H.,  {Ashok} N.~M.,  {Banerjee} D.~P.~K.,  {Valisa}
  P.,  {Milani} A.,  {Siviero} A.,  {Dallaporta} S.,    {Castellani} F.,  2011,
  \mnras, 410, L52

\bibitem[\protect\citeauthoryear{{Nelson}, {Chomiuk}, {Roy}, {Sokoloski},
  {Mukai}, {Krauss}, {Mioduszewski}, {Rupen} \& {Weston}}{{Nelson}
  et~al.}{2014}]{Nelson14}
{Nelson} T.,  {Chomiuk} L.,  {Roy} N.,  {Sokoloski} J.~L.,  {Mukai} K.,
  {Krauss} M.~I.,  {Mioduszewski} A.~J.,  {Rupen} M.~P.,    {Weston} J.,  2014,
  \apj, 785, 78

\bibitem[\protect\citeauthoryear{{Nelson}, {Donato}, {Mukai}, {Sokoloski} \&
  {Chomiuk}}{{Nelson} et~al.}{2012}]{Nelson12}
{Nelson} T.,  {Donato} D.,  {Mukai} K.,  {Sokoloski} J.,    {Chomiuk} L.,
  2012, \apj, 748, 43

\bibitem[\protect\citeauthoryear{{Ness}, {Osborne}, {Henze}, {Dobrotka},
  {Drake}, {Ribeiro}, {Starrfield}, {Kuulkers}, {Behar}, {Hernanz}, {Schwarz},
  {Page}, {Beardmore} \& {Bode}}{{Ness} et~al.}{2013}]{Ness13}
{Ness} J.-U.,  {Osborne} J.~P.,  {Henze} M.,  {Dobrotka} A.,  {Drake} J.~J.,
  {Ribeiro} V.~A.~R.~M.,  {Starrfield} S.,  {Kuulkers} E.,  {Behar} E.,
  {Hernanz} M.,  {Schwarz} G.,  {Page} K.~L.,  {Beardmore} A.~P.,    {Bode}
  M.~F.,  2013, \aap, 559, A50

\bibitem[\protect\citeauthoryear{{O'Brien}, {Bode}, {Porcas}, {Muxlow},
  {Eyres}, {Beswick}, {Garrington}, {Davis} \& {Evans}}{{O'Brien}
  et~al.}{2006}]{OBrien06}
{O'Brien} T.~J.,  {Bode} M.~F.,  {Porcas} R.~W.,  {Muxlow} T.~W.~B.,  {Eyres}
  S.~P.~S.,  {Beswick} R.~J.,  {Garrington} S.~T.,  {Davis} R.~J.,    {Evans}
  A.,  2006, \nat, 442, 279

\bibitem[\protect\citeauthoryear{{Orio}, {Rana}, {Page}, {Sokoloski} \&
  {Harrison}}{{Orio} et~al.}{2015}]{Orio15}
{Orio} M.,  {Rana} V.,  {Page} K.~L.,  {Sokoloski} J.,    {Harrison} F.,  2015,
  \mnras, 448, L35

\bibitem[\protect\citeauthoryear{{Page}, {Osborne}, {Wagner}, {Beardmore},
  {Shore}, {Starrfield} \& {Woodward}}{{Page} et~al.}{2013}]{Page13}
{Page} K.~L.,  {Osborne} J.~P.,  {Wagner} R.~M.,  {Beardmore} A.~P.,  {Shore}
  S.~N.,  {Starrfield} S.,    {Woodward} C.~E.,  2013, \apjl, 768, L26

\bibitem[\protect\citeauthoryear{{Ribeiro}, {Munari} \& {Valisa}}{{Ribeiro}
  et~al.}{2013}]{Ribeiro13}
{Ribeiro} V.~A.~R.~M.,  {Munari} U.,    {Valisa} P.,  2013, \apj, 768, 49

\bibitem[\protect\citeauthoryear{{Rupen}, {Mioduszewski} \&
  {Sokoloski}}{{Rupen} et~al.}{2008}]{Rupen08}
{Rupen} M.~P.,  {Mioduszewski} A.~J.,    {Sokoloski} J.~L.,  2008, \apj, 688,
  559

\bibitem[\protect\citeauthoryear{{Schaefer}, {Brummelaar}, {Gies},
  {Farrington}, {Kloppenborg}, {Chesneau}, 
  {Muirhead}}{{Schaefer} et~al.}{2014}]{Schaefer14}
{Schaefer} G.~H.,  {Brummelaar} T.~T.,  {Gies} D.~R.,  {Farrington} C.~D.,
  {Kloppenborg} B.,  {Chesneau} O.,  
  2014, \nat, 515, 234

\bibitem[\protect\citeauthoryear{{Seaquist} \& {Palimaka}}{{Seaquist} \&
  {Palimaka}}{1977}]{Seaquist77}
{Seaquist} E.~R.,  {Palimaka} J.,  1977, \apj, 217, 781

\bibitem[\protect\citeauthoryear{{Shepherd}}{{Shepherd}}{1997}]{Shepherd97}
{Shepherd} M.~C.,  1997, in {Hunt} G.,  {Payne} H.,  eds, Astronomical Data
  Analysis Software and Systems VI Vol.~125 of Astronomical Society of the
  Pacific Conference Series, {Difmap: an Interactive Program for Synthesis
  Imaging}.
p.~77

\bibitem[\protect\citeauthoryear{{Shore}, {Schwarz}, {De Gennaro Aquino},
  {Augusteijn}, {Walter}, {Starrfield} \& {Sion}}{{Shore}
  et~al.}{2013}]{Shore13}
{Shore} S.~N.,  {Schwarz} G.~J.,  {De Gennaro Aquino} I.,  {Augusteijn} T.,
  {Walter} F.~M.,  {Starrfield} S.,    {Sion} E.~M.,  2013, \aap, 549, A140

\bibitem[\protect\citeauthoryear{{Sokoloski}, {Luna}, {Mukai} \&
  {Kenyon}}{{Sokoloski} et~al.}{2006}]{Sokoloski06}
{Sokoloski} J.~L.,  {Luna} G.~J.~M.,  {Mukai} K.,    {Kenyon} S.~J.,  2006,
  \nat, 442, 276

\bibitem[\protect\citeauthoryear{{Sokoloski}, {Rupen} \&
  {Mioduszewski}}{{Sokoloski} et~al.}{2008}]{Sokoloski08}
{Sokoloski} J.~L.,  {Rupen} M.~P.,    {Mioduszewski} A.~J.,  2008, ApJ, 685,
  L137

\bibitem[\protect\citeauthoryear{{Sokolovsky}, {Korotkiy}, {Elenin} \&
  {Molotov}}{{Sokolovsky} et~al.}{2012}]{ATEL4061}
{Sokolovsky} K.,  {Korotkiy} S.,  {Elenin} L.,    {Molotov} I.,  2012, The
  Astronomer's Telegram, 4061, 1

\bibitem[\protect\citeauthoryear{{Sytov}, {Kaigorodov}, {Bisikalo}, {Kuznetsov}
  \& {Boyarchuk}}{{Sytov} et~al.}{2007}]{Sytov07}
{Sytov} A.~Y.,  {Kaigorodov} P.~V.,  {Bisikalo} D.~V.,  {Kuznetsov} O.~A.,
  {Boyarchuk} A.~A.,  2007, Astronomy Reports, 51, 836

\bibitem[\protect\citeauthoryear{{Thompson}}{{Thompson}}{2012}]{Stereo2012}
{Thompson} W., , 2012, STEREO HI1-B observations of Nova Sagittarii 2012,
  \url{http://stereo.gsfc.nasa.gov/~thompson/nova_sagittarii_2012}

\bibitem[\protect\citeauthoryear{{Walter}, {Battisti}, {Towers}, {Bond} \&
  {Stringfellow}}{{Walter} et~al.}{2012}]{Walter12}
{Walter} F.~M.,  {Battisti} A.,  {Towers} S.~E.,  {Bond} H.~E.,
  {Stringfellow} G.~S.,  2012, \pasp, 124, 1057

\bibitem[\protect\citeauthoryear{{Weiler}, {Panagia}, {Montes} \&
  {Sramek}}{{Weiler} et~al.}{2002}]{Weiler02}
{Weiler} K.~W.,  {Panagia} N.,  {Montes} M.~J.,    {Sramek} R.~A.,  2002,
  \araa, 40, 387

\bibitem[\protect\citeauthoryear{{Weston}, {Sokoloski}, {Metzger}, {Zheng},
  {Chomiuk}, {Krauss}, {Linford}, {Nelson}, {Mioduszewski}, {Rupen}, {Finzell}
  \& {Mukai}}{{Weston} et~al.}{2016}]{Weston16}
{Weston} J.~H.~S.,  {Sokoloski} J.~L.,  {Metzger} B.~D.,  {Zheng} Y.,
  {Chomiuk} L.,  {Krauss} M.~I.,  {Linford} J.~D.,  {Nelson} T.,
  {Mioduszewski} A.~J.,  {Rupen} M.~P.,  {Finzell} T.,    {Mukai} K.,  2016,
  \mnras, 457, 887

\bibitem[\protect\citeauthoryear{{Weston}, {Sokoloski}, {Zheng}, {Chomiuk},
  {Mioduszewski}, {Mukai}, {Rupen}, {Krauss}, {Roy} \& {Nelson}}{{Weston}
  et~al.}{2014}]{Weston13}
{Weston} J.~H.~S.,  {Sokoloski} J.~L.,  {Zheng} Y.,  {Chomiuk} L.,
  {Mioduszewski} A.,  {Mukai} K.,  {Rupen} M.~P.,  {Krauss} M.~I.,  {Roy} N.,
   {Nelson} T.,  2014, in {Woudt} P.~A.,  {Ribeiro} V.~A.~R.~M.,  eds, Stell
  Novae: Past and Future Decades Vol.~490 of Astronomical Society of the
  Pacific Conference Series, {Shocks and Ejecta Mass: Radio Observations of
  Nova V1723 Aql}.
p.~339

\bibitem[\protect\citeauthoryear{{Williams}}{{Williams}}{2012}]{Williams12}
{Williams} R.,  2012, \aj, 144, 98

\bibitem[\protect\citeauthoryear{{Wilms}, {Allen} \& {McCray}}{{Wilms}
  et~al.}{2000}]{Wilms00}
{Wilms} J.,  {Allen} A.,    {McCray} R.,  2000, \apj, 542, 914

\end{thebibliography}

\end{document}